%% file: main.tex
\documentclass[11pt,a4paper]{article}
\pdfoutput=1 

\newcommand{\Nb}{\bar{N}}
\newcommand{\gb}{{\bar g}}

\newcommand{\lam}{\lambda}
\newcommand{\muf}{\mu_F}

\newcommand{\g}{{\cal G}}

\def\LogmW1{{{\ln (1-\omega)}}}

\usepackage[table]{xcolor}
\usepackage{colortbl}
\RequirePackage{ifpdf} 
\usepackage{amsmath} 
\usepackage{mathtools}
\usepackage{cases}

\usepackage{jheppub}
\usepackage{pstricks}
\usepackage[final]{pdfpages} 
\usepackage{ifpdf} 
\usepackage{slashed}

\usepackage{color}
\usepackage{xcolor}
\definecolor{urlblue}{rgb}{0.2,0.4,0.7}
\definecolor{citegreen}{rgb}{0,0.6,0.2}
\definecolor{linkred}{rgb}{0.9,0.2,0.1}
\usepackage{hyperref}
\hypersetup{
colorlinks=true, citecolor=citegreen, linkcolor=blue, urlcolor=urlblue}

\usepackage{graphics}
\usepackage{etoolbox} 
\usepackage{fixmath}
\usepackage{psfrag}
\usepackage{bm}

\usepackage{notoccite} 

\usepackage{amsfonts}
\usepackage{autobreak}
\usepackage{marginnote}

\usepackage{tikz}
\usetikzlibrary{positioning,arrows}
\usetikzlibrary{decorations.pathmorphing}
\usetikzlibrary{decorations.markings}
\usetikzlibrary{shapes.geometric}
\tikzset{
    vector/.style={decorate, decoration={snake}, draw},
    provector/.style={decorate, decoration={snake,amplitude=2.5pt}, draw},
    antivector/.style={decorate, decoration={snake,amplitude=-2.5pt}, draw},
    fermion/.style={draw=black, postaction={decorate},decoration={markings,mark=at position .55 with {\arrow[draw=black]{>}}}},
    fermionbar/.style={draw=black, postaction={decorate},
                       decoration={markings,mark=at position .55 with {\arrow[draw=black]{<}}}},
    fermionnoarrow/.style={draw=black},
    gluon/.style={decorate, draw=black,decoration={coil,amplitude=4pt, segment length=5pt}},
    scalar/.style={dashed,draw=black, postaction={decorate},decoration={markings,mark=at position .55 with {\arrow[draw=black]{>}}}},
    scalarbar/.style={dashed,draw=black, postaction={decorate},decoration={markings,mark=at position .55 with {\arrow[draw=black]{<}}}},
    scalarnoarrow/.style={dashed,draw=black},
    electron/.style={draw=black, postaction={decorate},decoration={markings,mark=at position .55 with {\arrow[draw=black]{>}}}},
    bigvector/.style={decorate, decoration={snake,amplitude=4pt}, draw},
}

\def\g{\mathcal{\tilde{G}}}
\def\gm{\gamma}

\def\epm1{\frac{1}{\epsilon}}
\def\epm2{\frac{1}{\epsilon^{2}}}
\def\epm3{\frac{1}{\epsilon^{3}}}
\def\epm4{\frac{1}{\epsilon^{4}}}

\def\gm0{\gamma_{0}}
\def\gm1{\gamma_{1}}
\def\gm2{\gamma_{2}}
\def\gm3{\gamma_{3}}
\def\nn{\nonumber}

\def\nn{\nonumber\\}

\def\arXiv{{\fontfamily{qcr}\selectfont
arXiv}}

\title{Soft-virtual correction and threshold resummation for $n$-colorless particles to fourth order in QCD: Part I}

\author[a,b]{Taushif Ahmed,}
\author[c]{ Ajjath A. H.,}
\author[d]{Goutam Das,}
\author[c]{Pooja Mukherjee,}
\author[c]{V. Ravindran}
\author[c]{and Surabhi Tiwari}
\affiliation[a]{Max-Planck-Institut f\"ur Physik, Werner-Heisenberg-Institut, 80805 M\"unchen, Germany}
\affiliation[b]{Dipartimento di Fisica and Arnold-Regge Center, Universit\`a di Torino, 
\\ and INFN, Sezione di Torino, Via Pietro Giuria 1, I-10125 Torino, Italy}
\affiliation[c]{The Institute of Mathematical Sciences, HBNI, IV Cross Road, Taramani, Chennai 600113, India}
\affiliation[d]{Theoretische Physik 1, Naturwissenschaftlich-Technische Fakult{\"a}t, Universit{\"a}t Siegen,
Walter-Flex-Strasse 3, 57068 Siegen, Germany}
\emailAdd{taushif@mpp.mpg.de}
\emailAdd{ajjathah@imsc.res.in}
\emailAdd{goutam.das@uni-siegen.de}
\emailAdd{poojamukherjee@imsc.res.in}
\emailAdd{ravindra@imsc.res.in}
\emailAdd{surabhit@imsc.res.in}

\preprint{MPP-2020-51, SI-HEP-2020-02}

\abstract{
We present the general form of a universal soft-collinear distribution operator to compute the soft-virtual cross-section to next-to-next-to-next-to-next-to-leading order (N$^4$LO) in QCD for a process with any number of final state colorless particles in hadron colliders. By acting this universal operator on the pure virtual corrections, which need to be computed explicitly for a process, the soft-virtual cross-section can be obtained. 
The operator is constructed by exploiting the factorization and renormalization group evolution of amplitudes in QCD, and the universality of soft gluon contributions. We also provide the hard coefficient to perform the threshold resummation to next-to-next-to-next-to-leading logarithmic (N$^3$LL) accuracy. Furthermore, we present the approximate analytical results of the soft-virtual cross-sections at N$^4$LO and N$^3$LL for the Higgs boson production through gluon fusion and bottom quark annihilation, and also for the Drell-Yan production at the hadronic collider.}

\begin{document}
\allowdisplaybreaks[4]
\unitlength1cm
\maketitle 

\flushbottom

\input{intro.tex}

\input{sv-Xsection.tex}
\input{SCoperator.tex}

\input{resum.tex}
\input{concl.tex}

\appendix
\input{AppZ2ir.tex}

\input{AppSI.tex}
\input{AppDI.tex}

\input{AppSdI.tex}
\input{AppCRI.tex}
\input{N4LO_New.tex}

\input{N3LL.tex}

\bibliography{main} 
\bibliographystyle{JHEP}
 
\end{document}

%% file: intro.tex
\section{Introduction}
\label{sec:intro}

The post Higgs discovery era of high energy physics has witnessed a rapidly growing precision at the hadron colliders which in turn puts increasing strain on our ability to make reliable theoretical predictions. Multi-loop and multi-leg calculations are among the primary ladders to achieve the theoretically precise results for the processes currently produced in abundance at the large hadron collider (LHC). Over the past few decades, an abundant amount of results of physical observables have been computed to next-to-next-to-leading (NNLO) order in QCD for $2\rightarrow 2$ scattering\footnote{For summary of the recent state-of-the-art, see~\cite{Brooijmans:2020yij}.} and to next-to-next-to-next-to-leading (N$^3$LO) only for a few number of $2\rightarrow 1$ processes~\cite{Anastasiou:2015ema,Mistlberger:2018etf,Dulat:2018bfe,Duhr:2019kwi,Mondini:2019gid,Duhr:2020seh,Duhr:2020kzd}. Very recently, NNLO QCD correction to $2 \rightarrow 3$ scattering is achieved in ref.~\cite{Chawdhry:2019bji} and N$^3$LO correction to $2 \rightarrow 2$ process is obtained in ref.~\cite{Chen:2019lzz,Chen:2019fhs}.

The demand of high precision theoretical results for multi-leg final state colorless particles are of crying need due to its importance in determining many physical quantities, such as, self-coupling of the discovered Higgs boson. However, with the increase of loops and legs, the complexity of the computation grows so rapidly that often it becomes very difficult to compute a physical observable exactly. In this scenario, in absence of the exact fixed order computation, we could try to capture the dominant contributions to a physical observable by evaluating the quantity in some limit. In this article, we focus on such an approximation to the scattering cross-section where we capture only the contribution arising from soft region, which is called the soft-virtual (SV) approximation. When a system of large invariant mass is produced at the hadronic collision, generically the SV result turns out to be the dominant one. Since the partonic distribution function, $f_a(x)$, grows very rapidly with decreasing momentum fraction $x$, the cross-section gets enhanced when $x$ is very close to its minima. This, in turn, implies that the incoming partonic center-of-mass energy tends to be very close to the invariant mass of the final state particles, and the remaining energy can only produce some soft particles. Therefore, this limit is called the soft approximation. 

In this article, we focus on a generic $2 \rightarrow n$ scattering in hadronic collision with final state consisting of only colorless particles, such as gauge bosons, Higgs, leptonic pair in Drell-Yan. The soft gluon resummation for $2 \rightarrow n$ process in single particle inclusive kinematics is studied in ref.~\cite{Forslund:2020lnu}. In ref.~\cite{deFlorian:2012za}, the universal soft part to obtain the SV cross-section at NNLO is presented.  In ref.~\cite{Catani:2014uta}, an elegant approach was developed
to obtain the SV cross-section up to N$^3$LO with the help of  infrared subtraction operators and using universal soft gluon contributions.  The present article
provides an alternative approach to obtain the results of \cite{Catani:2014uta}. Moreover, by exploiting the factorization property and renormalization group evolution of QCD amplitude, we give an alternative prescription to compute the SV correction to the inclusive production cross-section to next-to-next-to-next-to-next-to-leading order (N$^4$LO) in QCD. With this \textit{universal} expression, the SV cross-section of any process of the aforementioned kind can be computed in an automated way once the virtual amplitudes for $2 \rightarrow n$ process become available and, therefore, provides a first estimate of the size of higher order corrections. 
Among the ingredients that enter into the computation of N$^4$LO SV cross-section, the complete result of the light-like cusp anomalous dimension at four loops becomes available recently~\cite{Korchemsky:1987wg,Moch:2004pa,Vogt:2004mw,Henn:2019swt,vonManteuffel:2020vjv}. The remaining universal quantities, namely, the virtual and soft anomalous dimensions, which are related to the collinear anomalous dimension through a relation conjectured in ref.~\cite{Ravindran:2004mb}, are also not fully available yet in its analytic form at four loops.

The prescription is based on refs.~\cite{Ravindran:2005vv,Ravindran:2006cg} where the framework is formulated for $2 \rightarrow 1$ process and subsequently applied to compute several observables under SV approximation.  In ref.~\cite{Ravindran:2005vv},
from the inclusive cross sections at NNLO for Drell-Yan (DY) and Higgs productions, the universal soft-collinear distribution 
was obtained which demonstrated the property of maximally non-abelian.  In other words, the soft-collinear distribution of DY is related to that of Higgs production by Casimir scaling, that is, by a ratio $C_F/C_A$.  Here, $C_F$ and $C_A$ are Casimirs
of SU$(n_c)$ in fundamental and adjoint representations, respectively.  Conjecturing that the Casimir scaling would hold
even at three loops, third order contribution to the soft-collinear distribution for DY was obtained simply from the
N$^3$LO result for the Higgs production computed in the SV approximation.  The SV cross-sections of the production of Drell-Yan pair~\cite{Ahmed:2014cla}, Higgs boson through bottom quark annihilation~\cite{Ahmed:2014cha}, Higgs in association with vector boson~\cite{Kumar:2014uwa}, pseudo-scalar Higgs boson~\cite{Ahmed:2015qda}, and massive spin-2~\cite{Das:2019bxi} are computed to N$^3$LO employing the prescription. In ref.~\cite{Catani:2014uta}, the N$^3$LO correction for DY was computed independently and it agreed with that of 
\cite{Ahmed:2014cla}, see also \cite{Li:2014bfa} for an explicit computation.   The methodology is also successfully extended to incorporate two final state colorless particles which is employed to compute the SV NNLO cross-section of the production of a pair of Higgs boson~\cite{H:2018hqz} and Higgs boson in association with $Z$ boson~\cite{Ahmed:2019udm,Ahmed:2019syo} through bottom quark annihilation. In order to incorporate the QED and mixed QCD$\oplus$QED corrections, this prescription is used to compute the corresponding soft contribution to NNLO in ref.~\cite{H:2019nsw} and N$^3$LO in ref.~\cite{Ajjath:2019vmf}. Furthermore, this prescription is also extended to compute the differential rapidity distribution for a $2 \rightarrow 1$ scattering process under SV approximation in ref.~\cite{Ravindran:2006bu} which is employed to calculate the corresponding quantity at N$^3$LO for the massive vector boson in ref.~\cite{Ravindran:2007sv} and, Drell-Yan pair and Higgs boson in refs.~\cite{Ahmed:2014uya,Ahmed:2014era}. The gluon jet function to N$^3$LO is computed in ref.~\cite{Banerjee:2018ozf} employing the formalism in the context of deep inelastic scattering. In addition, to compute the NNLO splitting functions in ${\cal N}=4$ supersymmetric Yang-Mills theory~\cite{Banerjee:2018yrn,Ahmed:2019hcp}, this prescription is employed extensively. The SV cross-section of the productions of Higgs boson in gluon fusion and Drell-Yan at N$^3$LO are computed in ref.~\cite{Anastasiou:2014vaa,Li:2014bfa} by explicitly evaluating the real emission Feynman diagrams in the soft limit.


For $n$-colorless particle production, the final state invariant mass can take any value from partonic threshold to the hadronic center of mass energy and thereby the hadronic threshold making resummation essential. Even, away from hadronic threshold, the resummation plays crucial role for sub-percent accuracy at the colliders. Note that the threshold resummation is important when average partonic center of mass energy is also small. The size of the resummed effect depends on how the partonic threshold is reached in the hadronic colliders. In particular, it is controlled by the shape of the parton distribution functions. For PDFs which peak at lower $x$ values, resummation is thus important far away from hadronic threshold, for example, in Higgs production.

SV part of the cross-section satisfies renormalization group on its own. The highest distributions can be completely predicted from the lower orders itself. The formalism on how to resum these large threshold logarithms has been developed decades ago from the seminal works by Stermann~\cite{Sterman:1986aj}, Catani and Trentedue~\cite{Catani:1989ne}. Threshold resummation is conveniently performed in the Mellin-$N$ space to avoid the convolution structure in the distribution space. In the Mellin-space the plus distributions translate into simple logarithms in the Mellin variable $N$. The threshold limit in distribution space $z\to 1$ gets translated into large-$N$ limit ($N \to \infty$) in Mellin space and the threshold enhanced terms can be organized in a \textit{resummed series} which determines the accuracy of the resummation. 

Threshold resummation has been widely used in the past decades to improve several processes in the Standard Model (SM), for example inclusive Higgs production in gluon fusion \cite{Catani:2003zt,Moch:2005ky,Catani:2014uta,Bonvini:2014joa,Bonvini:2016frm} and bottom quark annihilation \cite{H:2019dcl}, Drell-Yan invariant mass distribution \cite{Moch:2005ky,Idilbi:2006dg,Catani:2014uta,Eynck:2003fn,H.:2020ecd}, deep inelastic scattering \cite{Moch:2005ba,Das:2019btv} , vector boson production \cite{H.:2020ecd} are known to next-to-next-to-next-to-leading logarithmic (N$^3$LL) accuracy and found to improve the perturbative predictions. Several studies are done even beyond SM (BSM) scenario for example in inclusive pseudo-scalar production \cite{Ahmed:2016otz,Schmidt:2015cea,deFlorian:2007sr}, spin-2 production \cite{Das:2019bxi,das:rs}. 

In this article we present a novel approach on how to systematically perform threshold resummation to N$^3$LL for $n$-colorless particle production  at the hadron collider such as the LHC. Starting from the factorization at the threshold limit, we show how to extract the relevant ingredients, in particular, the Mellin-$N$ independent terms. 

We also provide the approximate results of the SV cross-sections at N$^4$LO and N$^3$LL explicitly for the Higgs boson production through gluon fusion and bottom quark annihilation, and for the Drell-Yan production. The approximation comes from the unavailability of the full four loop virtual matrix elements and the soft-collinear distributions. The \textit{goal} of this article is to present the formal methodology of computing threshold corrections for the inclusive production cross-section for  any generic process of $2 \rightarrow n$ kind at hadron collider. 

The article is organized as follows. In section.~\ref{sec:Xsection}, we introduce the notion of soft-virtual cross-section. The \textit{universal} soft-collinear operator to obtain the SV cross-section to N$^4$LO for a generic scattering of $2 \rightarrow n$ is presented in section~\ref{sec:phi}. The \textit{universal} soft-collinear operator that is needed to perform the threshold resummation to N$^3$LL is obtained in section~\ref{sec:resum}. After drawing the conclusion in section~\ref{sec:conclusion}, we present all the results in the appendix for a specific scale choice. With the \arXiv~submission of this article, we provide a Mathematica readable ancillary file containing all the results of the appendix.

%% file: sv-Xsection.tex
\section{Soft-virtual cross-section}
\label{sec:Xsection}
In this section, we develop the methodology to obtain soft gluon contributions to the inclusive production of $n$-number of 
colorless particles in hadron collisions within the framework of perturbative QCD. 
We use QCD improved parton model, namely, the factorization of collinear divergences from the partonic
reactions to all orders in strong coupling constant $\alpha_s=g_s^2/4 \pi$.  
We start by considering a general hadronic collision between two hadrons $H_{1,(2)}$ with momentum $P_{1(2)}$ that produces a final state composed of $n$-number of colorless particles $F_i(q_i)$
\begin{align}
\label{eq:had-collision}
    H_1(P_1) + H_2(P_2) \rightarrow \sum_{i=1}^n F_i(q_i)+X\,.
\end{align}
The quantity $X$ represents an inclusive hadronic state and $q_i$ stands for the momenta of corresponding colorless particle $F_i$. We denote the invariant mass square of the final state by $q^2$ which is related to the momentum $\{q_i\}$ through $q^2=\left(\sum_i q_i \right)^2$. The invariant mass distribution of the final state is given by 
\begin{align}
\label{eq:hadronic-Xsection}
    q^2{d \over dq^2}\sigma(S,q^2)=\sum_{a,b}\int dx_1 f_{a} (x_1,\mu_F^2) \int dx_2 {f}_{b}(x_2,\mu_F^2)  
q^2{d \over dq^2}\hat \sigma_{a b}\left(x_1,x_2,S,q^2,\mu_F^2\right)\,,
\end{align}
where $S$ and ${s}$ are the square of the hadronic and partonic center-of-mass energy, respectively. $x_{1,2}$ are fractions of momenta of incoming hadrons carried away by partons $a$ and $b$, and ${f}_{a(b)}$ are the parton distribution function (PDF) renormalized at the factorization scale $\mu_F$. The strong coupling constant at renormalization scale $\mu_R$ is denoted by $a_s(\mu_R^2)\equiv\alpha_s(\mu_R^2)/4\pi$. 
The mass-factorized partonic cross-section, $\hat \sigma_{ab}$, is related to the matrix element, ${\cal M}_{ab}$, through
\begin{align}
\label{eq:Xsection-matrix-reln}
    \hat \sigma_{ab}=\frac{1}{2 s} \int \left[ dPS_m\right] \left| \overline{\cal M}_{ab} \right|^2\,,
\end{align}
where $\left[ dPS_m\right]$ represents the $m$-particle phase space. In the above equation, 
the overline on the matrix element 
means that  the sum over all color and spin of the final states and the average over the same for initial 
states are done before performing the phase space integration. The partonic cross-section can be perturbatively expanded in powers of $a_s(\mu_R^2)$ as
\begin{align}
\label{eq:pert-expn-sigma}
    q^2{d \over dq^2}\hat \sigma_{ab}(x_1,x_2,S,q^2,\mu_F^2) = a_s^{\lambda}(\mu_R^2) \sum_{k=0}^{\infty} a_s^k(\mu_R^2) q^2{d \over dq^2}\hat \sigma_{ab}^{(k)}
(x_1,x_2,S,q^2,\mu_F^2,\mu_R^2)\,,
\end{align}
where ${\lambda}$ is determined by the leading order (LO) process.  The partonic cross section is $\mu_F$ dependent
and $\mu_R$ independent.  The $\mu_F$ dependence goes away at the hadronic level when they are folded with
appropriate parton distribution functions and the integration over $x_1,x_2$ and sum over all the partons are performed.  
In the following, we restrict ourselves to only those partonic processes where
the LO process can either be quark initiated with same flavor, $q\bar q \rightarrow \sum_i F_i$ or gluon initiated, 
$gg \rightarrow \sum_i F_i$ owing to the color conservation. This implies 
\begin{align}
\label{eq:LO-Xsection}
    \hat \sigma^{(0)}_{ab}=(\delta_{aq}\delta_{b\bar q}+\delta_{a\bar q}\delta_{bq}+\delta_{ag} \delta_{bg}) C_0 \delta(1-z) \,,
\end{align}
where all the dependence of partonic cross-section on kinematic invariants at LO is encapsulated in $C_0$. The variable $z$ is defined as $z \equiv q^2/{s}$. Beyond LO, there can be many other subprocesses that contribute towards the total cross section and in addition to having the $\delta(1-z)$ part, there are various other terms involving plus-distribution, logarithm and constant. The plus-distribution is defined through
\begin{align}
\label{eq:plus-dist}
\int_{0}^1 dz {\cal D}_j(z) g(z) = \int_{0}^1 dz \left(\frac{\log^j(1-z)}{(1-z)} \right)_+ g(z) \equiv 
	\int_{0}^1 dz \frac{\log^j(1-z)}{(1-z)} [g(z)-g(1)] \,.
\end{align}

In this article, we are interested in computing the cross-section in the soft limit which translates to $z \rightarrow 1$. This limit implies that the initial partonic center of mass energy is almost completely exhausted to produce the final state colorless particles $\{F_i\}$ and the small residual energy can only produce some soft partons. Since in this limit, only soft radiation can take place, it is called the soft-limit. In this scenario when all the emitted partons are soft, we can approximate the partonic cross-section by its threshold expansion
\begin{align}
\label{eq:thres-expand}
    \hat \sigma^{(k)}_{ab}=(\delta_{aq}\delta_{b\bar q}+\delta_{a\bar q}\delta_{bq}+\delta_{ag} \delta_{bg}) \hat \sigma^{(k),{\rm sv}}_{ab} + {\cal O}(1-z)^0\,.
\end{align}
The first term in the aforementioned threshold expansion,
the so-called soft-virtual term, only receives contributions from the gluon-gluon or quark-antiquark initial state. The SV cross-sections are linear combinations of the distributions $\delta(1-z)$ and ${\cal D}_j(z)$. The sub-leading terms in the expansion are called the hard contributions. The {\textit{goal}} of this article is to provide a prescription to compute the SV cross-section at fourth order in strong coupling constant i.e. $\hat \sigma^{(4),{\rm sv}}_{ab}$.

In the soft limit, it is well known that the square of the real emission partonic matrix elements factorizes into the hard and soft parts, and on top of that, the phase space splits into the corresponding parts. Consequently, by combining the soft part with the pure virtual contribution and the mass factorized counter terms, we obtain the infrared safe SV hadronic cross-section which reads 
\begin{align}
\label{eq:Xsection-SV-Hadron}
    \sigma^{\rm sv} =& \int \frac{d q^2}{q^2} 
\sum_{ab} \int dx_1 f_a(x_1,\mu_F) \int dx_2 f_{b} (x_2,\mu_F)  
\frac{1}{2  s} \prod_{i=1}^n \int d \phi(q_i) 
\nonumber\\
&
\times (2 \pi)^D \delta^D\Big(p_1+p_2-\sum_{i=1}^n q_i\Big)
\Delta_{ab}^{\rm sv} 
\left(\{p_j\cdot q_k\},q^2,z,\mu_F^2\right)\,.
\end{align}
We denote the momentum of the initial state partons by $p_1,p_2$. All the scalar products among the external momentum are concisely represented by $\{p_j\cdot q_k\}$. The $d\phi(q_i)$ represents the $n$-particle Lorentz invariant phase-space element of the colorless final state particle corresponding to the momenta $q_i$ 
\begin{align}
\label{eq:LIPS}
    d\phi(q_i) \equiv \frac{d^3 q_i}{(2\pi)^3 2 E_i}\,,
\end{align}
where $q_i=(E_i,\Vec{q}_i)$. The quantity $\Delta_{ab}^{\rm sv}$ is called the SV coefficient function which is defined through
\begin{eqnarray}
\label{eq:def-SV-Cof-function}
\frac{\Delta^{\rm sv}(z)}{z} 
&=&\left(\Gamma_{\rm sv}^T(z,\mu_F^2,\epsilon) \right)^{-1}\otimes 
{\hat \Delta^{\rm sv}(z,\epsilon) \over z}
\otimes \Gamma_{\rm sv}^{-1}(z,\mu_F^2,\epsilon) 
\end{eqnarray}
where
\begin{eqnarray}
{\hat \Delta^{\rm sv}(z,\epsilon)\over z} = \frac{s}{2\pi} \int \Big( \prod_{j=1}^m d \phi(k_j) \Big) d \phi(q) 
(2\pi)^D \delta^{(D)}\Big(p_1+p_2-q-\sum_{j=1}^m k_j\Big) 
|\overline{{\cal M}}^{\rm sv}|^2  
\end{eqnarray}
In the aforementioned matrix equation, the quantity $D$ is the space-time dimensions and $|{\cal M}^{\rm sv} \rangle$ is the matrix element of the partonic subprocess
\begin{align}
\label{eq:process-partonic-level}
    a(p_1)+b(p_2) \rightarrow \sum_{i=1}^n F_i(q_i) + \sum_{j=1}^m r_j(k_j)\,,
\end{align}
where $r_j$ represents the partons emitted through real emission with momenta $k_j$. Beyond leading order, the real emission partons start appearing and their total number, $m$, depends on the order of perturbation that we are looking into. In the SV approximation, all the real emissions are assumed to be soft i.e. $k_j \rightarrow 0$ which is represented through the superscript ${\rm SV}$ in the matrix element. The overhead line on the matrix element denotes the average over initial states. $\Gamma(z,\mu_F^2,\epsilon)$ are the mass factorization kernels which remove the initial state collinear singularities through renormalization of parton distribution functions. The subscript ${\rm sv}$ in $\Gamma$ indicates
that we keep only $\delta(1-z)$ and ${\cal D}_j$ in $\Gamma$.  In the following section, we will present the prescription to calculate the coefficient function to N$^4$LO in QCD for any $2\rightarrow n$ scattering process, which can be calculated order by order in perturbation theory:
\begin{align}
\label{eq:pert-expn-Delta}
    \Delta_{ab}^{\rm sv}(\{p_i\cdot q_k\},z,q^2,\mu_F^2) = a_s^{\lambda}(\mu_R^2) \sum_{k=0}^{\infty} a_s^k(\mu_R^2) 
~\Delta_{ab}^{(k), {\rm sv}}(\{p_i\cdot q_k\},z,q^2,\mu_F^2,\mu_R^2)\,.
\end{align}

%% file: SCoperator.tex
\subsection{Universal soft-collinear operator for SV cross-section}
\label{sec:phi}

In constructing the cross-section, the UV renormalized virtual contribution is combined with the real emission diagrams and mass factorization in order to get rid of all the underlying soft and collinear (IR) singularities in dimensional regularization. A UV renormalized virtual amplitude for the scattering of $2\rightarrow n$ particles in dimensional regularization can always be factorized into a part containing all the IR poles and a finite part as
%
\begin{align}
\label{eq:M-Mfin}
    {\cal M}_{ab,{\rm fin}}\left(\{p_j\},\{ q_k\},\mu_R^2\right)  = \lim_{\epsilon \rightarrow 0} Z_{ab,\rm IR}^{-1}(q^2,\mu_R^2,\epsilon) {\cal M}_{ab}\left(\{p_j\}, \{q_k\},\epsilon\right) \,.
\end{align}
In the aforementioned equation, we factorize the infrared poles at the renormalization scale. In principle, the scale at which we factorize could be different. Here ${\cal M}_{ab,{\rm fin}}$ represents the finite part of the pure virtual amplitude that depends on momentum, colors and spins of the external states. The Feynman diagrams that contribute to the aforementioned pure virtual amplitude have the same kinematics as the leading order process.   For processes such as the Higgs boson production through gluon fusion in heavy quark effective field theory, the coupling constant renormalization is insufficient to remove all the UV divergences. We need to renormalize the operator by multiplying an overall operator renormalization constant, $Z_{\rm OP}$. This is a property inherently associated with the operator and it should not be mix with the UV renormalization constants for the couplings present in the theory. For conserved operator, such as leptonic pair production in Drell-Yan, this quantity is identically one.

The factor $Z_{ab,\rm IR}(q^2,\mu_R^2,\epsilon)$ is a universal quantity encapsulating all the soft and collinear divergences~\cite{Catani:1998bh,Sterman:2002qn,Ravindran:2004mb,Moch:2005tm,Aybat:2006wq,Aybat:2006mz,Becher:2009cu,Gardi:2009qi,Almelid:2015jia,Almelid:2017qju} which appear as poles in dimensional regulator $\epsilon \equiv (D-4)$. It depends only on the external colored partons, and independent of the nature and numbers of the colorless particles. The suffix $ab$ is used to signify the fact that the external partons solely determine the IR behavior of the virtual matrix element. The universal structures of the infrared divergences to two-loops (except the single pole) were first depicted by Catani in ref.~\cite{Catani:1998bh} in terms of subtraction operators which was later formally proved in ref.~\cite{Sterman:2002qn}. For Sudakov form factors that involves two external partons, the explicit form of the single pole at two-loop was first revealed in ref.~\cite{Ravindran:2004mb} whose validity at three-loop was later established in ref.~\cite{Moch:2005tm}. In ref.~\cite{Becher:2009cu,Gardi:2009qi}, an all order conjecture on the form of infrared divergences for any generic process is made in terms of soft anomalous dimension matrix. For readers' convenience, we present the $Z_{ab,{\rm IR}}(q^2,\mu_R^2,\epsilon)$ to four loops in appendix~\ref{secA:Z2IR} as well as in the ancillary file.

Since the IR behavior of the pure virtual amplitude is completely universal and independent of the number of external colorless particles, the combined contributions from the real emission diagrams and mass factorization must also exhibit the same universality in order to get the finite cross-section. By employing this universality and imposing the constraint of the finiteness on the cross-section, we determine the universal contribution from the latter part to obtain the SV cross-section for any generic $2 \rightarrow n$ scattering process to N$^4$LO, which we now turn to.

By following the similar prescription for the $2 \rightarrow 1$ scattering process~\cite{Ravindran:2005vv,Ravindran:2006cg}, we propose that the SV partonic coefficient function in \eqref{eq:Xsection-SV-Hadron} can be written as a Mellin convolution of the pure virtual contribution ${\cal F}$ and soft-collinear distribution $\Phi$ as
\begin{align}
\label{eq:SV-master-formula}
    \Delta^{\rm sv}_{a\overline a} \left(\{p_j\cdot q_k\},z,q^2,\mu_F^2\right) =&  | {\cal M}^{(0)}_{a\overline a} |^2 |{\cal F}_{a\overline a}(\{p_j\cdot q_k\},q^2,\epsilon)|^2 \delta(1-z) 
 \nonumber\\  & 
    \otimes {\cal C} \exp\left({2\Phi_{a\overline a} \left(z,q^2,\epsilon\right)-2 {\cal C} \log \Gamma_{a\overline a}(z,\mu_F^2,\epsilon)}\right)\,.
\end{align}
The virtual contribution is captured through the form factor which is defined by the born factorized square matrix element as
\begin{align}
\label{eq:FF-defn}
    {\cal F}_{a\overline a} = 1+\sum_{k=1}^{\infty} a_s^k {\cal F}_{a\overline a}^{(k)} \equiv 1+\sum_{k=1}^{\infty} a_s^k \frac{\langle {\cal M}_{a\overline a}^{(0)}|{\cal M}_{a\overline a}^{(k)}\rangle}{\langle {\cal M}_{a\overline a}^{(0)}|{\cal M}_{a\overline a}^{(0)} \rangle}\,,
\end{align}
where ${\cal M}_{a\overline a}^{(k)}$ is the $k$-th order UV renormalized matrix element of the underlying partonic process $a(p_1)+\overline a(p_2) \rightarrow \sum_{i=1}^n F_i(q_i)$. The ``${\cal C}$ ordered exponential" of a distribution $g(z)$ containing $\delta(1-z)$ and ${\cal D}_j(z)$ is defined as
\begin{align}
\label{eq:Cordered-expoen}
    {\cal C}e^{g(z)} = \delta(1-z)+ \frac{1}{1!} g(z)+ \frac{1}{2!} \left(g \otimes g\right)(z) + \frac{1}{3!} \left(g \otimes g \otimes g\right)(z)+\cdots\,,
\end{align}
where $\otimes$ denotes the Mellin convolution. Soft divergences from the real radiation, captured by the quantity $\Phi_{a\overline a}$, and from the virtual diagrams, encapsulated through ${\cal F}_{a\overline a}$, cancel with each other. The final state collinear singularities are guaranteed to cancel upon summing over final states, as dictated by  Kinoshita-Lee-Nauenberg (KLN) theorem. The initial state collinear singularities can arise from both the virtual and real emission diagrams which respectively show up in $F_{a\overline a}$ and $\Phi_{a\overline a}$, and upon incorporating the mass factorization kernels, $\Gamma$, all of these cease to exist. Consequently, the SV cross-section in \eqref{eq:SV-master-formula} is free of all the divergences. Hence, by demanding the finiteness of the $\Delta^{\rm sv}_{a\overline a}$, we can determine the explicit form of the newly introduced soft-collinear distribution $\Phi_{a\overline a}$. 

The real radiation for $2 \rightarrow 1$ and $2 \rightarrow n$ scattering processes can only occur from the initial state partons, and hence, $\Phi_{a\overline a}$ for the latter process must be same as that of the former one with modified kinematic dependence, dictated by the overall momentum conservation. Therefore, by understanding the behavior of the Sudakov ($2 \rightarrow 1$) form factors through its evolution under the momentum transfer $q^2$, which leads to its exponentiation~\cite{Mueller:1979ih,Collins:1980ih,Sen:1981sd}, and the renormalization group evolution of the mass factorization kernels~\cite{Moch:2004pa,Vogt:2004mw} enables us to systematically fix the mathematical structure of $\Phi_{a\overline a}$. The evolution of the form factor is explicitly present to five loops order in QCD in refs.~\cite{Moch:2005tm,Ravindran:2005vv,Ravindran:2006cg,Ahmed:2017gyt}. The $\Phi_{a\overline a}$ for the Sudakov form factor is determined to NNLO in refs.~\cite{Ravindran:2005vv,Ravindran:2006cg} and at N$^3$LO in ref.~\cite{Ahmed:2014cla}. In this article, for the first time, we present the $\Phi_{a\overline a}$ to N$^4$LO for any generic scattering process $2 \rightarrow n$. One of the most salient features of the $\Phi_{a\overline a}$ is that it satisfies the maximally non-Abelian property:
\begin{align}
\label{eq:Max-Non-Abelian}
    \Phi_{gg} = \frac{C_A}{C_F} \Phi_{q\bar q}\,,
\end{align}
where $C_F=(n_c^2-1)/2n_c$ and $C_A=n_c$ are the Casimirs in fundamental and Adjoint representation of SU($n_c$) gauge group. This property essentially signifies its universal behavior. Moreover, it is independent of the external quark flavors, as expected from the infrared behavior of the scattering amplitudes. The aforementioned non-Abelian property is explicitly verified to NNLO in refs.~\cite{Ravindran:2005vv,Ravindran:2006cg} and in ref.~\cite{Ahmed:2014cla} it is conjectured to be valid even at N$^3$LO QCD which is verified through explicit computations in refs.~\cite{Li:2014bfa,Catani:2014uta}. The flavor dependence of the $\Phi_{a\bar a}$ was exploited in ref.~\cite{Ahmed:2014cha} in order to calculate the SV cross-section at N$^3$LO for the Higgs boson production in bottom quark annihilation. However, whether the validity of this wonderful property holds beyond N$^3$LO with generalized Casimir scaling~\cite{Moch:2018wjh}, that needs to be addressed in future.


In order to make it more transparent, we decompose the soft-collinear distribution and logarithm of the mass factorization kernel into singular (sing) and finite (fin) parts as
\begin{align}
\label{eq:phi-sing-fin}
    &\Phi_{a\overline a}=\Phi_{a\overline a,{\rm sing}}+\Phi_{a\overline a,{\rm fin}}\,,\nonumber\\ 
    &\log \Gamma_{a\overline a}=\log \Gamma_{a\overline a,{\rm sing}}+\log \Gamma_{a\overline a,{\rm fin}}\,,
\end{align}
and rewrite \eqref{eq:SV-master-formula} as
\begin{align}
\label{eq:SV-Master-Rewrite}
        \Delta^{\rm sv}_{a\overline a} \left(\{p_j\cdot q_k\},z,q^2,\mu_F^2\right) =&  | {\cal M}^{(0)}_{a\overline a} |^2 |{\cal F}_{a\overline a,{\rm fin}}(\{p_j\cdot q_k\},q^2,\mu_R^2)|^2 \delta(1-z) 
 \nonumber\\  & 
    \otimes {\cal C} \exp\left({2\Phi_{a\overline a,{\rm fin}} \left(z,q^2,\mu_R^2\right)-2 {\cal C} \log \Gamma_{a\overline a,{\rm fin}}(z,\mu_R^2,\mu_F^2)}\right) 
    \otimes \textbf{I}_{a\overline a}^{\rm sv} \,,
\end{align}
where
\begin{align}
\label{eq:Isv}
    \textbf{I}_{a\overline a}^{\rm sv} =& |Z_{a{\bar a},{\rm IR}}(q^2,\mu_R^2, \epsilon)|^2 \delta(1-z) 
    \nonumber\\&
    \otimes \; {\cal C} \exp\left({2\Phi_{a\overline a,{\rm sing}} \left(z,q^2,\mu_R^2,\epsilon\right)-2 {\cal C} \log \Gamma_{a\overline a,{\rm sing}}(z,\mu_R^2,\epsilon)}\right)\,.
\end{align}
The finiteness of the SV cross-section $\Delta^{\rm sv}_{a\overline a}$ demands that the quantity $\textbf{I}_{a\overline a}^{\rm sv}$ must be equal to $\delta(1-z)$. The finite combination of soft-collinear distribution and mass factorization kernel in \eqref{eq:SV-Master-Rewrite} is universal which we call as \textit{soft-collinear operator} and this is denoted through ${\textbf{S}}_{a\overline a}$:
\begin{align}
\label{eq:soft-collinear-operator}
    {\textbf{S}}_{a\overline a} (z,q^2,\mu_R^2,\mu_F^2) \equiv {\cal C} \exp\left({2\Phi_{a\overline a,{\rm fin}} \left(z,q^2,\mu_R^2\right)-2 {\cal C} \log \Gamma_{a\overline a,{\rm fin}}(z,\mu_R^2,\mu_F^2)}\right) \,.
\end{align}
%
Being a universal quantity, it can be used for any generic $2 \rightarrow n$ scattering process. As depicted through  \eqref{eq:thres-expand}, the subprocesses with initial state partons as either $q\bar q$ or $gg$ are the only ones which contribute to the SV cross-section, so we can rewrite the aforementioned equation \eqref{eq:SV-Master-Rewrite} as
\begin{align}
\label{eq:rewrite-SV-I}
    \Delta^{\rm sv}_{I} = |  {\cal M}^{(0)}_{I} |^2  |{\cal F}_{I,{\rm fin}}|^2 \delta(1-z) \otimes \textbf{S}_{I}\,,
\end{align}
where $I$ can be either $q$ or $g$ for $q\bar q$ and $gg$ initiated processes, respectively. We expand the SV coefficient function in powers of $a_s(\mu_R)$ as
\begin{align}
\label{eq:pert-expn-SV-cof}
    \Delta_I^{\rm sv}(\{p_j\cdot q_k\},z,q^2,\mu_F^2) = a_s^{\lambda}(\mu_R^2) \sum_{k=0}^{\infty} a_s^k(\mu_R^2) \Delta_{I}^{(k),{\rm sv}}(\{p_j\cdot q_k\},z,q^2,\mu_F^2,\mu_R^2)\,.
\end{align}
The generic result of the soft-collinear operator and the SV coefficient function in terms of the universal light-like cusp ($A^I$), eikonal ($f^I$) and virtual ($B^I$) anomalous dimensions up to N$^4$LO are presented in the appendix~\ref{secA:SCdist} and \ref{secA:SVDelta}, respectively. We also provide the generic result by keeping the scale dependence explicitly as ancillary file with the \arXiv~submission. In order to obtain the complete analytical result at four loops by employing the general result presented in this article, one needs the four loop contribution to all the aforementioned anomalous dimensions along with the pure virtual amplitude.

%
%
The anomalous dimensions are expanded in powers of $a_s(\mu_R^2)$ as
\begin{align}
\label{eq:AD-expand}
    X^I(\mu_R^2) = \sum_{j=1}^{\infty} a_s^j(\mu_R^2) X^I_j\,, 
\end{align}
where $X=A,B,f$. As a consequence of recent calculations, the light-like  cusp anomalous dimensions are available to four loops~\cite{Korchemsky:1987wg,Moch:2004pa,Vogt:2004mw,Henn:2019swt,vonManteuffel:2020vjv} in QCD. The eikonal and virtual anomalous dimensions to three loops can be extracted~\cite{Ravindran:2004mb,Moch:2005tm} from the quark and gluon collinear anomalous dimensions~\cite{Becher:2006mr,Becher:2009qa} through the conjecture~\cite{Ravindran:2004mb}
\begin{align}
\label{eq:conjecture}
    \gamma^I=2 B^I + f^I\,.
\end{align}
The partial results of the eikonal and virtual anomalous dimensions at four loop can be obtained from refs.~\cite{Davies:2016jie,vonManteuffel:2020vjv,Das:2019btv,Das:2020adl}.

To provide an explicit results, we present the new results of $\Delta_{I}^{\rm sv}$ for $2\rightarrow 1$ processes such as the Drell-Yan and the Higgs boson productions through gluon fusion as well as bottom quark annihilation at N$^4$LO in appendix \ref{secD:N4LO}. In order to achieve this, we make use of the explicit results of the recently computed four loop cusp anomalous dimension~\cite{Korchemsky:1987wg,Moch:2004pa,Vogt:2004mw,Henn:2019swt,vonManteuffel:2020vjv} along with the available results of the form factors for Drell-Yan and the Higgs boson productions. The latter quantities are partially available to fourth order in QCD~\cite{Henn:2016men,Lee:2016ixa,vonManteuffel:2016xki,Lee:2017mip,vonManteuffel:2019wbj,vonManteuffel:2020vjv}. The previously missing coefficients of $\delta(1-z)$ in $\Delta_{d,I}^{\rm sv}$ at N$^4$LO were primarily due to the missing ${\cal O}(\epsilon^0)$ results of the form factors and soft-collinear distribution at four loop. The partial results for the SV cross-section at N$^4$LO, namely the coefficients of ${\cal D}_2$ to ${\cal D}_7$ for $I=q,g,b$, were first computed in ref.~\cite{Ravindran:2006cg} and the ${\cal D}_1$ for $I=q,g$ in ref.~\cite{,Ahmed:2014cla} (See version 1 on~\arXiv). The full explicit results of the quark and gluon form factors corresponding to Drell-Yan and the Higgs boson production in gluon fusion are available at three loop to ${\cal O}(\epsilon^2)$ in ref.~\cite{,Gehrmann:2010ue,Gehrmann:2010tu}. The corresponding partial four loop form form factors are computed in several articles over the past decade~\cite{Henn:2016men,Lee:2016ixa,vonManteuffel:2016xki,Lee:2017mip,vonManteuffel:2019wbj,vonManteuffel:2020vjv}. In case of Higgs boson production through bottom quark annihilation, the three loop~\cite{Gehrmann:2014vha,Lee:2017mip} and partial four loop form factors are available in ref.~\cite{Lee:2017mip}. In addition, the one- and two-loop results are also needed to expand to ${\cal O}(\epsilon^5)$  and ${\cal O}(\epsilon^3)$, respectively. Similarly for the coefficient which contributes to the ${\cal O}(\epsilon^0)$ part of four loop form factor, the one-, two-, and three-loop results are needed to order ${\cal O}(\epsilon^6)$, ${\cal O}(\epsilon^4)$ and ${\cal O}(\epsilon^2)$, respectively. The four loop explicit results which we present in this article are still incomplete due to the unavailability of the full explicit results for form factors as well as soft contributions resulting from the real emission processes at four loop. Our findings of the coefficients of ${\cal D}_1$ for $I=q,g$ are consistent with the recent results computed in refs.~\cite{Das:2020adl}. The same quantity for $I=b$ is a new result that is presented for the first time in this article.


In ancillary file that is provided with the \arXiv~submission, the explicit expressions of all the anomalous dimensions including the QCD $\beta$-functions to three loops can be found. Since not all of these anomalous dimensions at four loop are fully present in the literature, we do not include the partial result in the ancillary file. In the following section, we focus on the soft gluon resummation for the $2 \rightarrow n$ process.
%


%% file: resum.tex
\section{Threshold resummation and its universal soft-collinear operator}
\label{sec:resum}
In the previous section, we have shown that the contributions resulting from 
soft gluon emissions in the production of colorless particles
in hadron collisions demonstrate the universal structure, namely, they are independent of
the hard process under study and they depend only on the type of incoming partons that is responsible for the
hard process. We have also shown that these contributions exponentiate to all orders in perturbation theory,
thanks to the factorization and renormalization group equation that they satisfy.  
The dominant contributions resulting from soft gluons are often studied in the
Mellin space by computing Mellin $N$-moment of partonic reactions in the large $N$ limit.  This approach
is advantageous due to absence of convolutions involving distributions.  Most importantly, as was
shown by the seminal works by Stermann~\cite{Sterman:1986aj}, Catani and Trentedue~\cite{Catani:1989ne},  it provides a 
systematic way of exponentiating large $\log N$
terms and of resumming $2 \beta_0 a_s(\mu_R^2) \log \Nb$ to all order in perturbation theory, where $\Nb \equiv N \exp(\gamma_E)$ and $\gamma_E$ is Euler-Mascheroni constant..    
In the following, we relate the process independent soft-collinear operator given in~\eqref{eq:soft-collinear-operator} to the well known
integral representation of the exponent in the Mellin space approach.  In addition, we determine
the process dependent part of the resummed result.   We decompose both $\Phi_{I,{\rm fin}}$ and
$\log\Gamma_{I,{\rm fin}}$ in \eqref{eq:phi-sing-fin} as
\begin{align}
\label{eq:dist-delt}
 &\Phi_{I,{\rm fin}}  =  \Phi_{I,{\cal D}} +  \Phi_{I,\delta} \,,\nonumber\\
     &\log \Gamma_{I,{\rm fin}} =\log \Gamma_{I,{\cal D}}+\log \Gamma_{I,{\delta}}\,,
\end{align}
where the quantities with subscript ${\cal D}$ contain only plus distributions ${\cal D}_j$ and
with the subscript $\delta$ contain only $\delta(1-z)$. 
The finite part of the mass factorization kernel $\Gamma_{I,{\rm fin}}$ contains 
$\log(\mu_F^2/\mu_R^2)$ resulting from  coupling constant renormalization.   
Substituting \eqref{eq:dist-delt} in \eqref{eq:SV-Master-Rewrite}, we obtain
\begin{align}
\label{eq:SVresum}
        \Delta^{\rm sv}_{I} \left(\{p_j\cdot q_k\},z,q^2,\muf^2\right) 
=
         C_{0}^I\left(\{p_j\cdot q_k\},q^2,\muf^2\right) \delta(1-z) \otimes {\cal C} \exp( \Phi^{res}_{I} (z,q^2,\mu_F^2))
    \otimes  \textbf{I}_{I}^{\rm sv}\,,
\end{align}
where the coefficient $C_0^I$ is given by 
\begin{eqnarray}
\label{eq:resc0}
 C_{0}^I\left(\{p_j\cdot q_k\},q^2,\muf^2\right)  
 &=& | {\cal M}^{(0)}_{I} |^2 |{\cal F}_{I,{\rm fin}}(\{p_j\cdot q_k\},q^2,\mu_R^2)|^2  S_{res,\delta}^I(q^2,\mu_R^2,\mu_F^2) 
\end{eqnarray}
with the soft-collinear operator for threshold resummation
\begin{eqnarray}
\label{eq:sdelta}
S_{res,\delta}^I(q^2,\mu_R^2,\mu_F^2) = \exp\left({2\Phi_{I,\delta}(q^2,\mu_R^2,\mu_F^2) -2  \log \Gamma_{I,{\delta}}}(\mu_F^2)  \right)\,.
\end{eqnarray}
The plus distributions ${\cal D}_i$ are contained in 
\begin{eqnarray}
\label{eq:resphi}
\Phi^{res}_{I}(z,q^2,\mu_F^2)  &=& 2 \Phi_{I, {\cal D}} (z,q^2)  - 2 {\cal C} \log \Gamma_{I, {\cal D}}(z,\mu_F^2)\,.
\end{eqnarray}
As it was shown in \cite{Ravindran:2005vv,Ravindran:2006cg} (see Eq.(44, 45, 46) of \cite{Ravindran:2006cg}) in the context of $2\rightarrow 1$ process,  
$\Phi^{res}_{I} $ takes the following form:
\begin{eqnarray}
\label{eq:phires}
\Phi^{res}_{I}(z,q^2,\mu_F^2)  &=&  
\Bigg( \frac{1}{1-z} \left\{ \int_{\muf^2}^{(1-z)^2 q^2} \frac{d\lam^2}{\lam^2} 2A^I(a_s(\lam^2)) + D^I(a_s((1-z)^2q^2)) \right\} \Bigg)_+ \,,
\end{eqnarray}
where the function $D^I$ are provided in the ancillary file.
It is convenient to perform the resummation in Mellin space. By taking the Mellin-$N$ moment of the \eqref{eq:SVresum}, we obtain
\begin{align}
\label{eq:res1}
\int_0^1 dz z^{N-1} \Delta^{\rm sv}_{I} \left(\{p_j\cdot q_k\},z,\muf^2\right) 
= C_{0}^I\left(\{p_j\cdot q_k\},q^2,\muf^2\right) \exp\left( \int_0^1 dz z^{N-1 }\Phi^{res}_{I}(z,q^2,\mu_F^2) \right)    \,.
\end{align}
The threshold limit in $z$ space gets translated to $N  \to \infty$ in the $N$ space. Following the method discussed in ref.~\cite{Catani:2003zt}, in the large $N$ limit 
\begin{eqnarray}
\label{eq:resGN}
  \int_0^1 dz z^{N-1 }\Phi^{res}_{I}(z,q^2,\mu_F^2)  &=&  \overline G_{0}^I(q^2,\mu_F^2) + 
\overline G_{\Nb}^I(q^2,\mu_F^2,\omega)\,,
\end{eqnarray}
where $\overline G_{0}^I$ is $N$ independent and $\overline G_{\Nb}^I$ satisfies the condition 
\begin{eqnarray}
\overline G_{\Nb}^I(q^2,\mu_F^2,\omega)|_{\Nb = 1} = 0\,.
\end{eqnarray}
Consequently, we get
\begin{align}
\label{eq:res2}
\int_0^1 dz z^{N-1} \Delta^{\rm sv}_{I} \left(\{p_j\cdot q_k\},z,\muf^2\right)
= \overline {g}_{0}^I\left(\{p_j\cdot q_k\},q^2,\muf^2\right) 
\exp\Big( \overline G_{\Nb}^I\left( q^2, \muf^2,\omega\right)\Big)\,.
\end{align}
The coefficient $\gb_{0}^I$ gets finite contribution from the form factor, soft-collinear distribution as well as terms arising from Mellin-space transformation of plus distributions, and it can be written as
\begin{align}
\label{eq:g0bexpanded}
\overline g_{0}^I (\{p_j\cdot q_k\},q^2,\muf^2) &= C_{0}^I\left(\{p_j\cdot q_k\},q^2,\muf^2\right) 
                                          ~\exp\left(\overline {G}_{0}^I(q^2,\muf^2)\right)\,.
\end{align}
The exponent ${\overline G}_{\Nb}^I$ can be organized as a resummed perturbation series in Mellin space,
\begin{align}
\label{eq:GNbexp}
 {\overline G}_{\Nb}^I(q^2,\mu_F^2,\omega) = \ln \Nb {\overline g}_{1}^I(\omega) + \sum_{i=0}^\infty a_s^i(\mu_R^2){\overline g}_{i+2}^I(\omega,q^2,\mu_F^2, \mu_R^2) \,,
\end{align}
where $\omega = 2\beta_0 a_s(\mu_R^2) \ln \Nb$. 
Similarly, we can expand $\overline g_{0}^I$ in powers of $a_s(\mu_R^2)$ as
\begin{eqnarray}
\label{eq:g0bexp}
\overline g_0^I(\{p_j\cdot q_k\},q^2,\muf^2) = \sum_{i=0}^\infty a_s^i(\mu_R^2) \overline g_{0,i}^I(\{p_j\cdot q_k\},q^2,
\muf^2,\mu_R^2)
\end{eqnarray}
The successive terms in the above series \eqref{eq:GNbexp} along with the corresponding terms in \eqref{eq:g0bexp}) define the resummed accuracy LL, NLL, NNLL, N$^3$LL and so on.


The explicit form in \eqref{eq:res2} along with its components as given in \eqref{eq:g0bexpanded} and \eqref{eq:GNbexp} are suitable for resumming large logarithms to all orders. For calculating the Mellin convolution of the SV partonic cross-section in \eqref{eq:res2} with \eqref{eq:g0bexpanded}, we need the process specific coefficient function $C^I_0$, and the universal quantities ${\overline G}^I_0$ and ${\overline G}^I_{\Nb}$. The process dependence of the former quantity comes from the process specific form factor. In appendix~\ref{secA:resCons}, we provide the general expressions of the universal quantities appearing in \eqref{eq:g0bexpanded} and \eqref{eq:GNbexp} which are required to perform the threshold resummation to N$^3$LL in QCD. In appendix~\ref{secE:N3LL}, we explicitly present the constants ${\overline{g}}^I_4$, as defined in \eqref{eq:GNbexp}, for the Drell-Yan, and Higgs boson production through gluon fusion as well as bottom quark annihilation.

%% file: concl.tex
\section{Conclusions and Outlook}
\label{sec:conclusion}
Higher order radiative corrections at hadronic colliders play very important role not only to
test the validity of the SM but also to constrain parameters of the beyond SM.  Among several observables, 
inclusive cross sections are often measured  experimentally and also are better understood theoretically. 
We restrict the discussion of this article to the inclusive production of $n$-colorless particles at hadron colliders, and
in particular, we focus to study the role of soft gluons both in fixed order as well as resummed framework.  

Often soft gluons
in the partonic reactions dominate, where
the center of mass energy of the partonic scattering is almost close to the invariant mass of the final state 
colorless particles.  In this article, we have studied these effects for fixed order predictions 
to fourth order in strong coupling constant, namely N$^4$LO level
and also the resummed soft gluon effects to third order in leading logarithmic accuracy, N$^3$LL for a generic $2 \rightarrow n$ scattering process. Specifically, we have explicitly presented the general structures of the soft-virtual cross-section at N$^4$LO and resummed result at N$^3$LL in QCD for any process of this kind. This is achieved by employing the collinear factorization of the inclusive cross section, 
the renormalization group invariance, universality of perturbative infrared structure of scattering amplitudes, and the process independence of the soft-collinear distribution. Furthermore, we have also explicitly given the soft-virtual cross-sections for the Drell-Yan, and Higgs boson productions through gluon fusion as well as bottom quark annihilation at N$^4$LO and N$^3$LL after incorporating the recent computations of the four loop form factors and anomalous dimensions.




The soft-collinear distribution
is found to be process independent as long as the partonic scattering  is inclusive in $n$ colorless particles.  
This is due to the fact that both hard part of square of the scattering amplitudes and 
the phase space for $n$ colorless state factorize from those of soft gluons.  The latter being process
independent can be obtained from simpler processes, namely, Drell-Yan for quark initiated process 
and Higgs productions for gluon initiated process.   
Using this fact along with the collinear factorization and the renormalization group invariance,
we have obtained most general
expression for the fixed order as well as resummed inclusive cross sections in terms of finite part of the
amplitudes correspond to $2 \rightarrow n$ colorless particles to fourth order in perturbative QCD . 
The latter is the only process dependent quantity that is needed for a given inclusive hadronic
reaction to obtain the soft gluon contribution to inclusive scattering cross section both in fixed order
as well as in resummed frameworks. The analytical results with explicit scale dependence are provided in the ancillary file with the \arXiv~submission.

\section*{Acknowledgements}
The algebraic computations have been done with the latest version of the symbolic manipulation system {\sc Form}~\cite{Vermaseren:2000nd,Ruijl:2017dtg}. The work of T.A. received funding from the European Research Council (ERC) under the European Union’s Horizon 2020 research and innovation programme, \textit{Novel structures in scattering amplitudes} (grant agreement No. 725110). The research of G.D. is supported by the DFG within the Collaborative Research Center TRR 257 (``{\it Particle Physics Phenomenology after the Higgs Discovery}'').

%% file: AppZ2ir.tex
\section{Infrared subtraction matrix for the amplitude}
\label{secA:Z2IR}
In this section, we present the constant $Z_{I,\rm IR}(q^2,\mu_R^2,\epsilon)$ 
(see \eqref{eq:M-Mfin}) 
that encapsulates the soft and collinear 
divergences of amplitude for $2 \rightarrow n$ colorless particles in dimensional regularization to four-loops in perturbative 
QCD.  Expanding the constant in powers of $a_s(\mu_R^2)$ as
\begin{align}
    Z_{I,\rm IR}(q^2,\mu_R^2,\epsilon) = 1 + \sum_{k=1}^{\infty} a_s^k(\mu_R^2)Z_{I,\rm IR}^{(k)}(q^2,\epsilon)
\end{align}
we present the results up to four loops by setting $\mu_R^2=q^2$. The result with explicit scale dependence can be found from the ancillary file supplied with the \arXiv~submission.
\begin{align}
\label{eq:Z2IR-Res}
\begin{autobreak}
Z_{I,{\rm IR}}^{(1)} =

        \frac{1}{\epsilon^2}   \bigg\{
          - 2 A^I_1
          \bigg\}

       + \frac{1}{\epsilon}   \bigg\{
           f^I_1
          + 2 B^I_1
          - \pi A^I_1 i
          \bigg\}\,,
\end{autobreak}\nonumber\\
\begin{autobreak}
 Z_{I,{\rm IR}}^{(2)} =

        \frac{1}{\epsilon^4}   \bigg\{
           2 (A^I_1)^2
          \bigg\}

       + \frac{1}{\epsilon^3}   \bigg\{
          - 2 A^I_1 f^I_1
          - 4 A^I_1 B^I_1
          - 3 \beta_{0} A^I_1
          + 2 \pi (A^I_1)^2 i
          \bigg\}

       + \frac{1}{\epsilon^2}   \bigg\{ \frac{1}{2} (f^I_1)^2
          + 2 B^I_1 f^I_1
          + 2 (B^I_1)^2
          - \frac{1}{2} A^I_2
          + \beta_{0} f^I_1
          + 2 \beta_{0} B^I_1
          - 3 \zeta_2 (A^I_1)^2
          - \pi A^I_1 f^I_1 i
          - 2 \pi A^I_1 B^I_1 i
          - \pi \beta_{0} A^I_1 i
          \bigg\}

       + \frac{1}{\epsilon}   \bigg\{
           \frac{1}{2} f^I_2
          + B^I_2
          - \frac{1}{2} \pi A^I_2 i
          \bigg\}\,,
\end{autobreak}\nonumber\\
\begin{autobreak}
    Z_{I,{\rm IR}}^{(3)} =

        \frac{1}{\epsilon^6}   \bigg\{
          - \frac{4}{3} (A^I_1)^3
          \bigg\}

       + \frac{1}{\epsilon^5}   \bigg\{
           2 (A^I_1)^2 f^I_1
          + 4 (A^I_1)^2 B^I_1
          + 6 \beta_{0} (A^I_1)^2
          - 2 \pi (A^I_1)^3 i
          \bigg\}

       + \frac{1}{\epsilon^4}   \bigg\{ - A^I_1 (f^I_1)^2
          - 4 A^I_1 B^I_1 f^I_1
          - 4 A^I_1 (B^I_1)^2
          + A^I_1 A^I_2
          - 5 \beta_{0} A^I_1 f^I_1
          - 10 \beta_{0} A^I_1 B^I_1
          - \frac{44}{9} \beta_{0}^2 A^I_1
          + 6 \zeta_2 (A^I_1)^3
          + 2 \pi (A^I_1)^2 f^I_1 i
          + 4 \pi (A^I_1)^2 B^I_1 i
          + 5 \pi \beta_{0} (A^I_1)^2 i
          \bigg\}

       + \frac{1}{\epsilon^3}   \bigg\{  \frac{1}{6} (f^I_1)^3
          + B^I_1 (f^I_1)^2
          + 2 (B^I_1)^2 f^I_1
          + \frac{4}{3} (B^I_1)^3
          - \frac{1}{2} A^I_2 f^I_1
          - A^I_2 B^I_1
          - A^I_1 f^I_2
          - 2 A^I_1 B^I_2
          - \frac{16}{9} \beta_{1} A^I_1
          + \beta_{0} (f^I_1)^2
          + 4 \beta_{0} B^I_1 f^I_1
          + 4 \beta_{0} (B^I_1)^2
          - \frac{10}{9} \beta_{0} A^I_2
          + \frac{4}{3} \beta_{0}^2 f^I_1
          + \frac{8}{3} \beta_{0}^2 B^I_1
          - 3 \zeta_2 (A^I_1)^2 f^I_1
          - 6 \zeta_2 (A^I_1)^2 B^I_1
          - 6 \zeta_2 \beta_{0} (A^I_1)^2
          - \frac{1}{2} \pi A^I_1 (f^I_1)^2 i
          - 2 \pi A^I_1 B^I_1 f^I_1 i
          - 2 \pi A^I_1 (B^I_1)^2 i
          + \frac{3}{2} \pi A^I_1 A^I_2 i
          - 2 \pi \beta_{0} A^I_1 f^I_1 i
          - 4 \pi \beta_{0} A^I_1 B^I_1 i
          - \frac{4}{3} \pi \beta_{0}^2 A^I_1 i
          + \pi \zeta_2 (A^I_1)^3 i
          \bigg\}

       + \frac{1}{\epsilon^2}   \bigg\{  \frac{1}{2} f^I_1 f^I_2
          + B^I_2 f^I_1
          + B^I_1 f^I_2
          + 2 B^I_1 B^I_2
          - \frac{2}{9} A^I_3
          + \frac{2}{3} \beta_{1} f^I_1
          + \frac{4}{3} \beta_{1} B^I_1
          + \frac{2}{3} \beta_{0} f^I_2
          + \frac{4}{3} \beta_{0} B^I_2
          - 3 \zeta_2 A^I_1 A^I_2
          - \frac{1}{2} \pi A^I_2 f^I_1 i
          - \pi A^I_2 B^I_1 i
          - \frac{1}{2} \pi A^I_1 f^I_2 i
          - \pi A^I_1 B^I_2 i
          - \frac{2}{3} \pi \beta_{1} A^I_1 i
          - \frac{2}{3} \pi \beta_{0} A^I_2 i
          \bigg\}

       + \frac{1}{\epsilon}   \bigg\{  \frac{1}{3} f^I_3
          + \frac{2}{3} B^I_3
          - \frac{1}{3} \pi A^I_3 i
          \bigg\}\,,
\end{autobreak}\nonumber\\
\begin{autobreak}
    Z_{I,{\rm IR}}^{(4)} =

        \frac{1}{\epsilon^8}   \bigg\{  \frac{2}{3} (A^I_1)^4
          \bigg\}

       + \frac{1}{\epsilon^7}   \bigg\{ - \frac{4}{3} (A^I_1)^3 f^I_1
          - \frac{8}{3} (A^I_1)^3 B^I_1
          - 6 \beta_{0} (A^I_1)^3
          + \frac{4}{3} \pi (A^I_1)^4 i
          \bigg\}

       + \frac{1}{\epsilon^6}   \bigg\{  (A^I_1)^2 (f^I_1)^2
          + 4 (A^I_1)^2 B^I_1 f^I_1
          + 4 (A^I_1)^2 (B^I_1)^2
          - (A^I_1)^2 A^I_2
          + 8 \beta_{0} (A^I_1)^2 f^I_1
          + 16 \beta_{0} (A^I_1)^2 B^I_1
          + \frac{257}{18} \beta_{0}^2 (A^I_1)^2
          - 6 \zeta_2 (A^I_1)^4
          - 2 \pi (A^I_1)^3 f^I_1 i
          - 4 \pi (A^I_1)^3 B^I_1 i
          - 8 \pi \beta_{0} (A^I_1)^3 i
          \bigg\}

       + \frac{1}{\epsilon^5}   \bigg\{ - \frac{1}{3} A^I_1 (f^I_1)^3
          - 2 A^I_1 B^I_1 (f^I_1)^2
          - 4 A^I_1 (B^I_1)^2 f^I_1
          - \frac{8}{3} A^I_1 (B^I_1)^3
          + A^I_1 A^I_2 f^I_1
          + 2 A^I_1 A^I_2 B^I_1
          + (A^I_1)^2 f^I_2
          + 2 (A^I_1)^2 B^I_2
          + \frac{32}{9} \beta_{1} (A^I_1)^2
          - \frac{7}{2} \beta_{0} A^I_1 (f^I_1)^2
          - 14 \beta_{0} A^I_1 B^I_1 f^I_1
          - 14 \beta_{0} A^I_1 (B^I_1)^2
          + \frac{67}{18} \beta_{0} A^I_1 A^I_2
          - \frac{95}{9} \beta_{0}^2 A^I_1 f^I_1
          - \frac{190}{9} \beta_{0}^2 A^I_1 B^I_1
          - \frac{25}{3} \beta_{0}^3 A^I_1
          + 6 \zeta_2 (A^I_1)^3 f^I_1
          + 12 \zeta_2 (A^I_1)^3 B^I_1
          + 21 \zeta_2 \beta_{0} (A^I_1)^3
          + \pi (A^I_1)^2 (f^I_1)^2 i
          + 4 \pi (A^I_1)^2 B^I_1 f^I_1 i
          + 4 \pi (A^I_1)^2 (B^I_1)^2 i
          - 2 \pi (A^I_1)^2 A^I_2 i
          + 7 \pi \beta_{0} (A^I_1)^2 f^I_1 i
          + 14 \pi \beta_{0} (A^I_1)^2 B^I_1 i
          + \frac{95}{9} \pi \beta_{0}^2 (A^I_1)^2 i
          - 2 \pi \zeta_2 (A^I_1)^4 i
          \bigg\}

       + \frac{1}{\epsilon^4}   \bigg\{  \frac{1}{24} (f^I_1)^4
          + \frac{1}{3} B^I_1 (f^I_1)^3
          + (B^I_1)^2 (f^I_1)^2
          + \frac{4}{3} (B^I_1)^3 f^I_1
          + \frac{2}{3} (B^I_1)^4
          - \frac{1}{4} A^I_2 (f^I_1)^2
          - A^I_2 B^I_1 f^I_1
          - A^I_2 (B^I_1)^2
          + \frac{1}{8} (A^I_2)^2
          - A^I_1 f^I_1 f^I_2
          - 2 A^I_1 B^I_2 f^I_1
          - 2 A^I_1 B^I_1 f^I_2
          - 4 A^I_1 B^I_1 B^I_2
          + \frac{4}{9} A^I_1 A^I_3
          - \frac{28}{9} \beta_{1} A^I_1 f^I_1
          - \frac{56}{9} \beta_{1} A^I_1 B^I_1
          + \frac{1}{2} \beta_{0} (f^I_1)^3
          + 3 \beta_{0} B^I_1 (f^I_1)^2
          + 6 \beta_{0} (B^I_1)^2 f^I_1
          + 4 \beta_{0} (B^I_1)^3
          - \frac{29}{18} \beta_{0} A^I_2 f^I_1
          - \frac{29}{9} \beta_{0} A^I_2 B^I_1
          - \frac{17}{6} \beta_{0} A^I_1 f^I_2
          - \frac{17}{3} \beta_{0} A^I_1 B^I_2
          - \frac{20}{3} \beta_{0} \beta_{1} A^I_1
          + \frac{11}{6} \beta_{0}^2 (f^I_1)^2
          + \frac{22}{3} \beta_{0}^2 B^I_1 f^I_1
          + \frac{22}{3} \beta_{0}^2 (B^I_1)^2
          - \frac{13}{6} \beta_{0}^2 A^I_2
          + 2 \beta_{0}^3 f^I_1
          + 4 \beta_{0}^3 B^I_1
          - \frac{3}{2} \zeta_2 (A^I_1)^2 (f^I_1)^2
          - 6 \zeta_2 (A^I_1)^2 B^I_1 f^I_1
          - 6 \zeta_2 (A^I_1)^2 (B^I_1)^2
          + \frac{15}{2} \zeta_2 (A^I_1)^2 A^I_2
          - 9 \zeta_2 \beta_{0} (A^I_1)^2 f^I_1
          - 18 \zeta_2 \beta_{0} (A^I_1)^2 B^I_1
          - 11 \zeta_2 \beta_{0}^2 (A^I_1)^2
          + \frac{3}{2} \zeta_2^2 (A^I_1)^4
          - \frac{1}{6} \pi A^I_1 (f^I_1)^3 i
          - \pi A^I_1 B^I_1 (f^I_1)^2 i
          - 2 \pi A^I_1 (B^I_1)^2 f^I_1 i
          - \frac{4}{3} \pi A^I_1 (B^I_1)^3 i
          + \frac{3}{2} \pi A^I_1 A^I_2 f^I_1 i
          + 3 \pi A^I_1 A^I_2 B^I_1 i
          + \pi (A^I_1)^2 f^I_2 i
          + 2 \pi (A^I_1)^2 B^I_2 i
          + \frac{28}{9} \pi \beta_{1} (A^I_1)^2 i
          - \frac{3}{2} \pi \beta_{0} A^I_1 (f^I_1)^2 i
          - 6 \pi \beta_{0} A^I_1 B^I_1 f^I_1 i
          - 6 \pi \beta_{0} A^I_1 (B^I_1)^2 i
          + \frac{40}{9} \pi \beta_{0} A^I_1 A^I_2 i
          - \frac{11}{3} \pi \beta_{0}^2 A^I_1 f^I_1 i
          - \frac{22}{3} \pi \beta_{0}^2 A^I_1 B^I_1 i
          - 2 \pi \beta_{0}^3 A^I_1 i
          + \pi \zeta_2 (A^I_1)^3 f^I_1 i
          + 2 \pi \zeta_2 (A^I_1)^3 B^I_1 i
          + 3 \pi \zeta_2 \beta_{0} (A^I_1)^3 i
          \bigg\}

       + \frac{1}{\epsilon^3}   \bigg\{  \frac{1}{4} (f^I_1)^2 f^I_2
          + \frac{1}{2} B^I_2 (f^I_1)^2
          + B^I_1 f^I_1 f^I_2
          + 2 B^I_1 B^I_2 f^I_1
          + (B^I_1)^2 f^I_2
          + 2 (B^I_1)^2 B^I_2
          - \frac{2}{9} A^I_3 f^I_1
          - \frac{4}{9} A^I_3 B^I_1
          - \frac{1}{4} A^I_2 f^I_2
          - \frac{1}{2} A^I_2 B^I_2
          - \frac{2}{3} A^I_1 f^I_3
          - \frac{4}{3} A^I_1 B^I_3
          - \frac{5}{4} \beta_{2} A^I_1
          + \frac{2}{3} \beta_{1} (f^I_1)^2
          + \frac{8}{3} \beta_{1} B^I_1 f^I_1
          + \frac{8}{3} \beta_{1} (B^I_1)^2
          - \frac{3}{4} \beta_{1} A^I_2
          + \frac{7}{6} \beta_{0} f^I_1 f^I_2
          + \frac{7}{3} \beta_{0} B^I_2 f^I_1
          + \frac{7}{3} \beta_{0} B^I_1 f^I_2
          + \frac{14}{3} \beta_{0} B^I_1 B^I_2
          - \frac{7}{12} \beta_{0} A^I_3
          + 2 \beta_{0} \beta_{1} f^I_1
          + 4 \beta_{0} \beta_{1} B^I_1
          + \beta_{0}^2 f^I_2
          + 2 \beta_{0}^2 B^I_2
          - 3 \zeta_2 A^I_1 A^I_2 f^I_1
          - 6 \zeta_2 A^I_1 A^I_2 B^I_1
          - \frac{3}{2} \zeta_2 (A^I_1)^2 f^I_2
          - 3 \zeta_2 (A^I_1)^2 B^I_2
          - 4 \zeta_2 \beta_{1} (A^I_1)^2
          - 7 \zeta_2 \beta_{0} A^I_1 A^I_2
          - \frac{1}{4} \pi A^I_2 (f^I_1)^2 i
          - \pi A^I_2 B^I_1 f^I_1 i
          - \pi A^I_2 (B^I_1)^2 i
          + \frac{1}{4} \pi (A^I_2)^2 i
          - \frac{1}{2} \pi A^I_1 f^I_1 f^I_2 i
          - \pi A^I_1 B^I_2 f^I_1 i
          - \pi A^I_1 B^I_1 f^I_2 i
          - 2 \pi A^I_1 B^I_1 B^I_2 i
          + \frac{8}{9} \pi A^I_1 A^I_3 i
          - \frac{4}{3} \pi \beta_{1} A^I_1 f^I_1 i
          - \frac{8}{3} \pi \beta_{1} A^I_1 B^I_1 i
          - \frac{7}{6} \pi \beta_{0} A^I_2 f^I_1 i
          - \frac{7}{3} \pi \beta_{0} A^I_2 B^I_1 i
          - \frac{7}{6} \pi \beta_{0} A^I_1 f^I_2 i
          - \frac{7}{3} \pi \beta_{0} A^I_1 B^I_2 i
          - 2 \pi \beta_{0} \beta_{1} A^I_1 i
          - \pi \beta_{0}^2 A^I_2 i
          + \frac{3}{2} \pi \zeta_2 (A^I_1)^2 A^I_2 i
          \bigg\}

       + \frac{1}{\epsilon^2}   \bigg\{  \frac{1}{8} (f^I_2)^2
          + \frac{1}{3} f^I_1 f^I_3
          + \frac{2}{3} B^I_3 f^I_1
          + \frac{1}{2} B^I_2 f^I_2
          + \frac{1}{2} (B^I_2)^2
          + \frac{2}{3} B^I_1 f^I_3
          + \frac{4}{3} B^I_1 B^I_3
          - \frac{1}{8} A^I_4
          + \frac{1}{2} \beta_{2} f^I_1
          + \beta_{2} B^I_1
          + \frac{1}{2} \beta_{1} f^I_2
          + \beta_{1} B^I_2
          + \frac{1}{2} \beta_{0} f^I_3
          + \beta_{0} B^I_3
          - \frac{3}{4} \zeta_2 (A^I_2)^2
          - 2 \zeta_2 A^I_1 A^I_3
          - \frac{1}{3} \pi A^I_3 f^I_1 i
          - \frac{2}{3} \pi A^I_3 B^I_1 i
          - \frac{1}{4} \pi A^I_2 f^I_2 i
          - \frac{1}{2} \pi A^I_2 B^I_2 i
          - \frac{1}{3} \pi A^I_1 f^I_3 i
          - \frac{2}{3} \pi A^I_1 B^I_3 i
          - \frac{1}{2} \pi \beta_{2} A^I_1 i
          - \frac{1}{2} \pi \beta_{1} A^I_2 i
          - \frac{1}{2} \pi \beta_{0} A^I_3 i
          \bigg\}

       + \frac{1}{\epsilon}   \bigg\{  \frac{1}{4} f^I_4
          + \frac{1}{2} B^I_4
          - \frac{1}{4} \pi A^I_4 i
          \bigg\}\,.
\end{autobreak}
\end{align}
The anomalous dimensions that appear in the aforementioned results are expanded following the \eqref{eq:AD-expand}. These including the coefficients of QCD $\beta$-functions are provided explicitly in the ancillary files to three loops.

%% file: AppSI.tex
\section{Soft-collinear distribution for SV cross-section}
\label{secA:SCdist}
In this section, we present soft-collinear distribution  ${\textbf{S}}_I$, as defined in \eqref{eq:rewrite-SV-I},
in powers of $a_s(\mu_R^2)$ up to four loops.
Expanding the quantity in powers of $a_s$ as
\begin{align}
    {\textbf{S}}_I(z,q^2,\mu_R^2,\mu_F^2) = \delta(1-z)+ \sum_{i=1}^{\infty} a_s^i(\mu_R^2) \sum_{J=\delta,{\cal D}} {\textbf{S}}^{(i)}_{I,J}(z,q^2,\mu_R^2,\mu_F^2)\,J\,,
\end{align}
we present the results for $\mu_R^2=\mu_F^2=q^2$. The $\delta(1-z)$ is represented by $\delta$ in the aforementioned equation. The result with explicit scale dependence can be found from the ancillary file supplied with the \arXiv~submission.

The symbols ${\tilde{\cal G}}^{I,j}_k$ that appear in the aforementioned soft-collinear distributions are provided explicitly in the ancillary files with the \arXiv~submission.

%% file: AppDI.tex
\section{Soft-virtual partonic cross-section}
\label{secA:SVDelta}
We expand the $\Delta^{\rm sv}_I$ in \eqref{eq:rewrite-SV-I} in powers of $a_s(\mu_R^2)$ through
\begin{align}
\label{eqA:Sof-Coll-Expand}
    \Delta^{\rm sv}_I\left(\{p_j\cdot q_k\},z,q^2,\mu_F^2\right) &= \delta(1-z) |{\cal M}^{(0)}_{I, {\rm fin}}|^2 \nonumber\\
    &+  \sum_{i=1}^{\infty} a_s^{i}(\mu_R^2) \sum_{J=\delta,{\cal D}}\Delta^{{\rm sv},(i)}_{I,J}\left(\{p_j\cdot q_k\},z,q^2,\mu_F^2,\mu_R^2\right) J\,.
\end{align}
We denote the $\delta(1-z)$ by $\delta$. Here, we present $\Delta^{\rm sv}_I$ to fourth order for the specific scale choice $a_s(\mu_F^2)=a_s(\mu_R^2)=q^2$. 
The results with explicit dependence on $\mu_R$ and $\mu_F$ are provided in the ancillary files supplied with the \arXiv~submission.


In the aforementioned equations, we define
\begin{align}
    {\cal M}^{(m,n)}_{I,{\rm fin}} \equiv {\rm Real} \left( \langle {\cal M}^{(m)}_I|{\cal M}^{(n)}_I \rangle_{\rm fin} \right)
\end{align}
where, $|{\cal M}^{(n)}_I \rangle$ is the UV  renormalized pure virtual amplitude at $n$-th order in $a_s$ as introduced in \eqref{eq:M-Mfin}.

%% file: AppSdI.tex
\section{Soft-collinear distribution for threshold resummation}
\label{APP:Sdelta}
The universal soft-collinear operator, defined in \eqref{eq:sdelta}, that is required to obtain the resummed cross-section in $z$-space is expanded in powers of $a_s(\mu_R^2)$ as
\begin{align}
    S_{res,\delta}^I(q^2,\mu_R^2,\mu_F^2) = 1 + \sum_{i=1}^{\infty} a_s^i(\mu_R^2) S_{res,\delta}^{I,(i)}(q^2,\mu_R^2,\mu_F^2)\,.
\end{align}
We present the results for $q^2=\mu_R^2=\mu_F^2$ below to fourth order in coupling constant. The result with explicit scale dependence can be obtained from the ancillary files provided with the \arXiv~submission.
\begin{align}
\begin{autobreak}
   {\cal S}^{I,(1)}_{res,\delta} =
       2 \g^{I,(1)}_1\,,
\end{autobreak}\nonumber\\
\begin{autobreak}
   {\cal S}^{I(2)}_{res,\delta} =
       \g^{I,(1)}_2
       + 2 (\g^{I,(1)}_1)^2
       + 2 \beta_{0} \g^{I,(2)}_1\,,
\end{autobreak}\nonumber\\
\begin{autobreak}
   {\cal S}^{I(3)}_{res,\delta} =
       \frac{2}{3} \g^{I,(1)}_3
       + 2 \g^{I,(1)}_1 \g^{I,(1)}_2
       + \frac{4}{3} (\g^{I,(1)}_1)^3
       + \frac{4}{3} \beta_{1} \g^{I,(2)}_1
       + \frac{4}{3} \beta_{0} \g^{I,(2)}_2
       + 4 \beta_{0} \g^{I,(1)}_1 \g^{I,(2)}_1
       + \frac{8}{3} \beta_{0}^2 \g^{I,(3)}_1\,,
\end{autobreak}\nonumber\\
\begin{autobreak}
   {\cal S}^{I(4)}_{res,\delta} =
       \frac{1}{2} \g^{I,(1)}_4
       + \frac{1}{2} (\g^{I,(1)}_2)^2
       + \frac{4}{3} \g^{I,(1)}_1 \g^{I,(1)}_3
       + 2 (\g^{I,(1)}_1)^2 \g^{I,(1)}_2
       + \frac{2}{3} (\g^{I,(1)}_1)^4
       + \beta_{2} \g^{I,(2)}_1
       + \beta_{1} \g^{I,(2)}_2
       + \frac{8}{3} \beta_{1} \g^{I,(1)}_1 \g^{I,(2)}_1
       + \beta_{0} \g^{I,(2)}_3
       + 2 \beta_{0} \g^{I,(2)}_1 \g^{I,(1)}_2
       + \frac{8}{3} \beta_{0} \g^{I,(1)}_1 \g^{I,(2)}_2
       + 4 \beta_{0} (\g^{I,(1)}_1)^2 \g^{I,(2)}_1
       + 4 \beta_{0} \beta_{1} \g^{I,(3)}_1
       + 2 \beta_{0}^2 \g^{I,(3)}_2
       + 2 \beta_{0}^2 (\g^{I,(2)}_1)^2
       + \frac{16}{3} \beta_{0}^2 \g^{I,(1)}_1 \g^{I,(3)}_1
       + 4 \beta_{0}^3 \g^{I,(4)}_1\,.
\end{autobreak}
\end{align}

%% file: AppCRI.tex
\section{Universal quantities for resummation}
\label{secA:resCons}

By expanding the $\exp\left (\overline G_0^I(q^2,\mu_R^2)\right)$, as defined in \eqref{eq:g0bexpanded}, in powers of $a_s^i(\mu_R^2)$ as
\begin{align}
\exp\left (\overline G_0^I(q^2,\mu_R^2)\right) = 1+\sum_{i=1}^{\infty} a_s^i(\mu_R^2) {\tilde {\overline G}}^I_{0,i}(q^2,\mu_R^2)\,,
\end{align}
we present the general results to fourth order in QCD for $\mu_R^2=q^2$. The complete result containing the explicit scale dependence is provided with the \arXiv~submission.
\begin{align}
\begin{autobreak}
{\tilde {\overline G}}^I_{0,1} =
2 \zeta_2 A^I_1\,,
\end{autobreak}\nonumber\\
\begin{autobreak}
{\tilde {\overline G}}^I_{0,2} =
\frac{8}{3} \zeta_3 \beta_{0} A^I_1
       + 2 \zeta_2 A^I_2
       + 2 \zeta_2 \beta_{0} f^I_1
       + 2 \zeta_2^2 (A^I_1)^2\,,
\end{autobreak}\nonumber\\
\begin{autobreak}
{\tilde {\overline G}}^I_{0,3} =
\frac{8}{3} \zeta_3 \beta_{1} A^I_1
       + \frac{16}{3} \zeta_3 \beta_{0} A^I_2
       + \frac{16}{3} \zeta_3 \beta_{0}^2 f^I_1
       + 2 \zeta_2 A^I_3
       + 2 \zeta_2 \beta_{1} f^I_1
       + 4 \zeta_2 \beta_{0} f^I_2
       + 8 \zeta_2 \beta_{0}^2 \g^{I,(1)}_1
       + \frac{16}{3} \zeta_2 \zeta_3 \beta_{0} (A^I_1)^2
       + 4 \zeta_2^2 A^I_1 A^I_2
       + 4 \zeta_2^2 \beta_{0} A^I_1 f^I_1
       + \frac{36}{5} \zeta_2^2 \beta_{0}^2 A^I_1
       + \frac{4}{3} \zeta_2^3 (A^I_1)^3\,,
\end{autobreak}\nonumber\\
\begin{autobreak}
{\tilde {\overline G}}^I_{0,4} =
\frac{192}{5} \zeta_5 \beta_{0}^3 A^I_1
       + \frac{8}{3} \zeta_3 \beta_{2} A^I_1
       + \frac{16}{3} \zeta_3 \beta_{1} A^I_2
       + 8 \zeta_3 \beta_{0} A^I_3
       + \frac{40}{3} \zeta_3 \beta_{0} \beta_{1} f^I_1
       + 16 \zeta_3 \beta_{0}^2 f^I_2
       + 32 \zeta_3 \beta_{0}^3 \g^{I,(1)}_1
       + \frac{32}{9} \zeta_3^2 \beta_{0}^2 (A^I_1)^2
       + 2 \zeta_2 A^I_4
       + 2 \zeta_2 \beta_{2} f^I_1
       + 4 \zeta_2 \beta_{1} f^I_2
       + 6 \zeta_2 \beta_{0} f^I_3
       + 20 \zeta_2 \beta_{0} \beta_{1} \g^{I,(1)}_1
       + 12 \zeta_2 \beta_{0}^2 \g^{I,(1)}_2
       + 24 \zeta_2 \beta_{0}^3 \g^{I,(2)}_1
       + \frac{16}{3} \zeta_2 \zeta_3 \beta_{1} (A^I_1)^2
       + 16 \zeta_2 \zeta_3 \beta_{0} A^I_1 A^I_2
       + 16 \zeta_2 \zeta_3 \beta_{0}^2 A^I_1 f^I_1
       + 32 \zeta_2 \zeta_3 \beta_{0}^3 A^I_1
       + 2 \zeta_2^2 (A^I_2)^2
       + 4 \zeta_2^2 A^I_1 A^I_3
       + 4 \zeta_2^2 \beta_{1} A^I_1 f^I_1
       + 4 \zeta_2^2 \beta_{0} A^I_2 f^I_1
       + 8 \zeta_2^2 \beta_{0} A^I_1 f^I_2
       + 18 \zeta_2^2 \beta_{0} \beta_{1} A^I_1
       + 2 \zeta_2^2 \beta_{0}^2 (f^I_1)^2
       + \frac{108}{5} \zeta_2^2 \beta_{0}^2 A^I_2
       + 16 \zeta_2^2 \beta_{0}^2 \g^{I,(1)}_1 A^I_1
       + \frac{108}{5} \zeta_2^2 \beta_{0}^3 f^I_1
       + \frac{16}{3} \zeta_2^2 \zeta_3 \beta_{0} (A^I_1)^3
       + 4 \zeta_2^3 (A^I_1)^2 A^I_2
       + 4 \zeta_2^3 \beta_{0} (A^I_1)^2 f^I_1
       + \frac{72}{5} \zeta_2^3 \beta_{0}^2 (A^I_1)^2
       + \frac{2}{3} \zeta_2^4 (A^I_1)^4\,.
\end{autobreak}
\end{align}
In the context of threshold resummation in Mellin space, the other universal quantity ${\overline G}_{\Nb}^I(q^2,\mu_F^2)$ is also expanded in powers of $a_s(\mu_R^2)$ in \eqref{eq:GNbexp}. The general results of the expansion coefficients in terms of universal quantities are presented to N$^3$LL accuracy in the ancillary file supplied with the \arXiv~submission. 

%% file: N4LO_New.tex
\section{Explicit results at N$^4$LO for Drell-Yan and Higgs boson production}
\label{secD:N4LO}
Below we present the results of $\Delta^{\rm sv}_{I}$ for Drell-Yan ($I=q$) and the Higgs boson production via gluon fusion ($I=g$) as well as bottom quark annihilation ($I=b$) at fourth order in strong coupling constant with the scale choice as $\mu_F^2=\mu_R^2=q^2$. We provide only the newly computed parts. The coefficients of ${\cal D}_2$ to ${\cal D}_7$ can be found in ref.~\cite{Ravindran:2006cg} for all the three processes. The ${\cal D}_1$ parts for $I=q,g$ were presented in the version 1 of the ref.~\cite{Ahmed:2014cla}. The coefficients of ${\cal D}_0$ for Drell-Yan and the Higgs boson production in gluon fusion are consistent with the results presented in ref.~\cite{Das:2020adl} after we make use of the available results of soft~\cite{Das:2019btv} and collinear~\cite{vonManteuffel:2020vjv} anomalous dimensions. To obtain the soft anomalous dimension for the gluon from that of quark~\cite{Das:2019btv}, we use the generalized Casimir scaling. The ${\cal D}_0$ part of the Higgs boson production through bottom quark annihilation differs from that of Drell-Yan only because of the Yukawa coupling. For readers' convenience, we explicitly present the new results including the ${\cal D}_0$ part for $I=b$. The results with explicit dependence on $\mu_R$ and $\mu_F$ are provided up to N$^4$LO in the ancillary files supplied with the \arXiv~submission. 

\begin{align}
\begin{autobreak}
   \Delta_{q}^{\rm sv,(4)} =
         \delta(1-z) \bigg\{\frac{1}{2} \g^{q,(1)}_4
          + \frac{d^{abcd}_F d^{abcd}_A}{N_F}   \bigg[ 
            3520 \zeta_2 \zeta_5
          + 128 \zeta_2 \zeta_3
          - 1152 \zeta_2 \zeta_3^2
          - 384 \zeta_2^2
          - \frac{23808}{35} \zeta_2^4
          + 2 \chi^{q}_{10}
          \bigg] 

       +  n_f   \bigg[ 
          - \frac{2}{3} \g^{q,(2)}_3
          + \frac{d^{abcd}_F d^{abcd}_F}{N_F}  \bigg(
            \frac{6380}{3}
          - 2480 \zeta_7
          + \frac{190196}{27} \zeta_5
          - \frac{26828}{27} \zeta_3
          + \frac{1360}{9} \zeta_3^2
          - \frac{43132}{9} \zeta_2
          - \frac{1856}{3} \zeta_2 \zeta_5
          - \frac{584}{3} \zeta_2 \zeta_3
          + \frac{49664}{45} \zeta_2^2
          - \frac{7904}{15} \zeta_2^2 \zeta_3
          + \frac{83240}{189} \zeta_2^3 \bigg)\bigg] 

       +  C_A   \bigg[ 
            \frac{11}{3} \g^{q,(2)}_3
          \bigg] 

       +  C_F n_f n_{fv} N_4    \bigg[ 
          - \frac{784}{9}
          + \frac{880}{9} \zeta_5
          - 56 \zeta_3
          + \frac{352}{3} \zeta_3^2
          - \frac{1112}{9} \zeta_2
          - \frac{112}{3} \zeta_2 \zeta_3
          + \frac{152}{9} \zeta_2^2
          + \frac{2816}{135} \zeta_2^3
          \bigg] 

       +  C_F n_f^3   \bigg[ 
            \frac{65633}{972}
          + \frac{676}{27} \zeta_5
          + \frac{7568}{729} \zeta_3
          - \frac{24932}{243} \zeta_2
          - \frac{1496}{81} \zeta_2 \zeta_3
          - \frac{7502}{405} \zeta_2^2
          \bigg] 

       +  C_F C_A n_{fv} N_4    \bigg[ 
          - \frac{3740}{3}
          - \frac{16544}{9} \zeta_5
          - \frac{2068}{9} \zeta_3
          + \frac{7568}{3} \zeta_3^2
          - \frac{5720}{3} \zeta_2
          - 660 \zeta_2 \zeta_3
          + \frac{4312}{15} \zeta_2^2
          + \frac{214016}{315} \zeta_2^3
          + \frac{1}{4} \chi^{q}_{4}
          \bigg] 

       +  C_F C_A n_f^2   \bigg[ - \frac{667210703}{419904}
          - \frac{2704}{27} \zeta_5
          + \frac{1247135}{1458} \zeta_3
          + \frac{136}{9} \zeta_3^2
          + \frac{38753131}{17496} \zeta_2
          - \frac{11576}{27} \zeta_2 \zeta_3
          + \frac{93769}{270} \zeta_2^2
          - \frac{5018}{315} \zeta_2^3
          \bigg] 

       +  C_F C_A^2 n_f   \bigg[ 
          - \frac{7253202277}{52488}
          - \frac{422}{9} \zeta_7
          + \frac{962812}{45} \zeta_5
          + \frac{147966065}{1458} \zeta_3
          - \frac{130948}{9} \zeta_3^2
          - \frac{170544238}{2187} \zeta_2
          - \frac{14672}{9} \zeta_2 \zeta_5
          + \frac{283990}{27} \zeta_2 \zeta_3
          + \frac{480169}{180} \zeta_2^2
          - \frac{13312}{45} \zeta_2^2 \zeta_3
          - \frac{1198}{35} \zeta_2^3
          + 2 \chi^{q}_{1}
          \bigg] 

       +  C_F C_A^3   \bigg[ \frac{29551452589}{104976}
          + \frac{2321}{9} \zeta_7
          - \frac{280943}{5} \zeta_5
          - \frac{183199867}{729} \zeta_3
          + 45001 \zeta_3^2
          + \frac{1493033303}{8748} \zeta_2
          + \frac{104456}{9} \zeta_2 \zeta_5
          - \frac{1479776}{27} \zeta_2 \zeta_3
          - 48 \zeta_2 \zeta_3^2
          - \frac{6097471}{405} \zeta_2^2
          + \frac{168256}{45} \zeta_2^2 \zeta_3
          + \frac{115049}{210} \zeta_2^3
          - \frac{20032}{35} \zeta_2^4
          + 2 \chi^{q}_{6}
          \bigg] 

       +  C_F^2 n_{fv} N_4    \bigg[ 
            \frac{6668}{3}
          - \frac{17084}{3} \zeta_7
          + \frac{76936}{27} \zeta_5
          + \frac{9680}{27} \zeta_3
          - \frac{9184}{9} \zeta_3^2
          + \frac{21124}{9} \zeta_2
          - \frac{2144}{3} \zeta_2 \zeta_5
          + 564 \zeta_2 \zeta_3
          - \frac{11648}{45} \zeta_2^2
          - \frac{1568}{3} \zeta_2^2 \zeta_3
          + \frac{4588}{27} \zeta_2^3
          + \frac{1}{4} \chi^{q}_{5}
          \bigg] 

       +  C_F^2 n_f^2   \bigg[ 
          - \frac{1194071}{2916}
          - \frac{53848}{27} \zeta_5
          - \frac{895846}{729} \zeta_3
          + \frac{27452}{27} \zeta_3^2
          + \frac{290338}{729} \zeta_2
          + \frac{182032}{81} \zeta_2 \zeta_3
          - \frac{108077}{810} \zeta_2^2
          - \frac{111512}{945} \zeta_2^3
          \bigg] 

       +  C_F^2 C_A n_f   \bigg[ \frac{161137649299}{944784}
          - 3444 \zeta_7
          - \frac{1123193}{243} \zeta_5
          - \frac{5006700101}{26244} \zeta_3
          + \frac{13810316}{729} \zeta_3^2
          + \frac{1735457753}{17496} \zeta_2
          - \frac{161672}{135} \zeta_2 \zeta_5
          - \frac{3295558}{81} \zeta_2 \zeta_3
          - \frac{10521971}{2916} \zeta_2^2
          + \frac{2436236}{405} \zeta_2^2 \zeta_3
          + \frac{820672}{315} \zeta_2^3
          + 2 \chi^{q}_{2}
          \bigg] 

       +  C_F^2 C_A^2   \bigg[ 
          - \frac{139605518111}{236196}
          - \frac{14368}{45} \zeta_{5,3}
          + \frac{225010}{9} \zeta_7
          + \frac{273937351}{2430} \zeta_5
          + \frac{31130595887}{52488} \zeta_3
          - \frac{56606}{9} \zeta_3 \zeta_5
          - \frac{59169646}{729} \zeta_3^2
          - \frac{37574761847}{104976} \zeta_2
          + \frac{634556}{135} \zeta_2 \zeta_5
          + \frac{155957506}{729} \zeta_2 \zeta_3
          - \frac{95108}{27} \zeta_2 \zeta_3^2
          + \frac{365582077}{14580} \zeta_2^2
          - \frac{14441606}{405} \zeta_2^2 \zeta_3
          + \frac{5597168}{945} \zeta_2^3
          + \frac{433849}{375} \zeta_2^4
          + 2 \chi^{q}_{7}
          \bigg] 

       +  C_F^3 n_f   \bigg[ 
          - \frac{7723623865}{209952}
          + \frac{4134164}{63} \zeta_7
          + \frac{2345956}{81} \zeta_5
          + \frac{20886044}{243} \zeta_3
          - \frac{2305220}{81} \zeta_3^2
          - \frac{12082445}{324} \zeta_2
          - \frac{139808}{3} \zeta_2 \zeta_5
          + \frac{830656}{81} \zeta_2 \zeta_3
          + \frac{2225602}{1215} \zeta_2^2
          - \frac{232048}{27} \zeta_2^2 \zeta_3
          + \frac{2846752}{2835} \zeta_2^3
          + 2 \chi^{q}_{3}
          \bigg] 

       +  C_F^3 C_A   \bigg[ 
            \frac{154126124135}{419904}
          - \frac{46048}{15} \zeta_{5,3}
          - \frac{9267206}{21} \zeta_7
          - \frac{262246151}{810} \zeta_5
          - \frac{1517196365}{5832} \zeta_3
          + \frac{288728}{5} \zeta_3 \zeta_5
          + \frac{11475955}{81} \zeta_3^2
          + \frac{1096193101}{3888} \zeta_2
          + \frac{11855032}{45} \zeta_2 \zeta_5
          - \frac{13567742}{81} \zeta_2 \zeta_3
          - \frac{423232}{9} \zeta_2 \zeta_3^2
          + \frac{4113199}{972} \zeta_2^2
          + \frac{5844584}{135} \zeta_2^2 \zeta_3
          - \frac{22022516}{567} \zeta_2^3
          - \frac{12244}{1125} \zeta_2^4
          + 2 \chi^{q}_{8}
          \bigg] 
          
       +  C_F^4   \bigg[ 
          - \frac{6246665}{64}
          + \frac{10304}{3} \zeta_{5,3}
          + \frac{329480}{7} \zeta_7
          + \frac{3101464}{15} \zeta_5
          - 144091 \zeta_3
          + \frac{19003904}{45} \zeta_3 \zeta_5
          - \frac{797552}{27} \zeta_3^2
          - \frac{1672459}{24} \zeta_2
          - \frac{21408}{5} \zeta_2 \zeta_5
          + \frac{422672}{9} \zeta_2 \zeta_3
          - \frac{3502208}{27} \zeta_2 \zeta_3^2
          - \frac{64097}{3} \zeta_2^2
          + \frac{59392}{15} \zeta_2^2 \zeta_3
          + \frac{10588864}{315} \zeta_2^3
          - \frac{16309024}{525} \zeta_2^4
          + 2 \chi^{q}_{9}
          \bigg] 
       \bigg\}\,,

\end{autobreak}
\end{align}

\begin{align}
\begin{autobreak}

   \Delta_{g}^{\rm sv,(4)} =

         \delta(1-z)   \bigg\{

          \frac{1}{2} \g^{g,(1)}_4

       +  \frac{d^{abcd}_A d^{abcd}_A}{N_A}   \bigg[ 
            3520 \zeta_2 \zeta_5
          + 128 \zeta_2 \zeta_3
          - 1152 \zeta_2 \zeta_3^2
          - 384 \zeta_2^2
          - \frac{23808}{35} \zeta_2^4
          + 2 \chi^{g}_{7}
          \bigg] 

       +  n_f   \bigg[ 
          - \frac{2}{3} \g^{g,(2)}_3
          + \frac{d^{abcd}_F d^{abcd}_A}{N_A}   \bigg(
            768 \zeta_2^2
          - 1280 \zeta_2 \zeta_5
          - 256 \zeta_2 \zeta_3
          + 2 \chi^{g}_{5}
          \bigg)
          \bigg] 

       +  n_f^2  \frac{d^{abcd}_F d^{abcd}_F}{N_A}   \bigg[ 
          - \frac{18016}{9}
          - 1920 \zeta_5
          + 3040 \zeta_3
          + 1024 \zeta_3^2
          + \frac{768}{5} \zeta_2^2
          \bigg] 

       +  C_A   \bigg[ 
            \frac{11}{3} \g^{g,(2)}_3
          \bigg] 

       +  C_A n_f^3   \bigg[ 
          - \frac{943703}{2916}
          + \frac{20}{27} \zeta_5
          - \frac{99964}{729} \zeta_3
          + \frac{191272}{729} \zeta_2
          + \frac{4360}{81} \zeta_2 \zeta_3
          + \frac{19802}{405} \zeta_2^2
          \bigg] 

       +  C_A^2 n_f^2   \bigg[ 
            \frac{3373102453}{419904}
          - \frac{88522}{27} \zeta_5
          + \frac{236569}{324} \zeta_3
          + \frac{30896}{27} \zeta_3^2
          - \frac{57010715}{17496} \zeta_2
          + \frac{56981}{27} \zeta_2 \zeta_3
          - \frac{244133}{180} \zeta_2^2
          - \frac{96146}{945} \zeta_2^3
          \bigg] 

       +  C_A^3 n_f   \bigg[ 
            \frac{8163906247}{236196}
          + \frac{4312342}{63} \zeta_7
          + \frac{63040172}{1215} \zeta_5
          - \frac{18110071}{13122} \zeta_3
          - \frac{12681814}{729} \zeta_3^2
          + \frac{43967867}{4374} \zeta_2
          - \frac{6509288}{135} \zeta_2 \zeta_5
          - \frac{7706462}{243} \zeta_2 \zeta_3
          + \frac{147411413}{14580} \zeta_2^2
          - \frac{1271644}{405} \zeta_2^2 \zeta_3
          + \frac{4148362}{945} \zeta_2^3
          + 2 \chi^{g}_{1}
          \bigg] 

       +  C_A^4   \bigg[ \frac{24881127343}{944784}
          + \frac{2048}{45} \zeta_{5,3}
          - \frac{7284211}{21} \zeta_7
          - \frac{43313873}{243} \zeta_5
          - \frac{477562055}{6561} \zeta_3
          + \frac{21319426}{45} \zeta_3 \zeta_5
          + \frac{101023100}{729} \zeta_3^2
          - \frac{198676295}{26244} \zeta_2
          + \frac{36923084}{135} \zeta_2 \zeta_5
          + \frac{67082140}{729} \zeta_2 \zeta_3
          - \frac{4868308}{27} \zeta_2 \zeta_3^2
          - \frac{209748319}{7290} \zeta_2^2
          + \frac{7031266}{405} \zeta_2^2 \zeta_3
          - \frac{29530507}{1890} \zeta_2^3
          - \frac{240117439}{7875} \zeta_2^4
          + 2 \chi^{g}_{6}
          \bigg] 

       +  C_F n_f^3   \bigg[ 
          - \frac{233953}{486}
          + \frac{1280}{27} \zeta_5
          + \frac{8120}{27} \zeta_3
          + \frac{752}{3} \zeta_2
          - \frac{1280}{9} \zeta_2 \zeta_3
          + \frac{224}{45} \zeta_2^2
          \bigg] 

       +  C_F C_A n_f^2   \bigg[ 
            \frac{49991435}{5832}
          - \frac{47144}{27} \zeta_5
          - \frac{791588}{243} \zeta_3
          - \frac{4508}{9} \zeta_3^2
          - \frac{116471}{81} \zeta_2
          + 144 \zeta_2 \zeta_3
          - \frac{78659}{270} \zeta_2^2
          - \frac{24872}{945} \zeta_2^3
          \bigg] 

       +  C_F C_A^2 n_f   \bigg[ 
          - \frac{495711665}{34992}
          + \frac{57488}{9} \zeta_7
          + \frac{2933116}{405} \zeta_5
          + \frac{2787854}{729} \zeta_3
          - \frac{518272}{81} \zeta_3^2
          - \frac{159851}{972} \zeta_2
          - \frac{4960}{9} \zeta_2 \zeta_5
          + \frac{45302}{81} \zeta_2 \zeta_3
          + \frac{1288613}{540} \zeta_2^2
          + \frac{73168}{45} \zeta_2^2 \zeta_3
          + \frac{28256}{63} \zeta_2^3
          + 2 \chi^{g}_{2}
          \bigg] 

       +  C_F^2 n_f^2   \bigg[ 
            \frac{11401}{27}
          + \frac{7840}{3} \zeta_5
          - \frac{10088}{3} \zeta_3
          + \frac{1792}{3} \zeta_3^2
          - \frac{100}{9} \zeta_2
          + \frac{64}{3} \zeta_2 \zeta_3
          - \frac{424}{15} \zeta_2^2
          + \frac{27392}{945} \zeta_2^3
          \bigg] 

       +  C_F^2 C_A n_f   \bigg[ 
            \frac{153625}{81}
          - \frac{35312}{3} \zeta_7
          + \frac{59456}{9} \zeta_5
          + \frac{222467}{27} \zeta_3
          - \frac{7648}{3} \zeta_3^2
          + \frac{1450}{9} \zeta_2
          - 960 \zeta_2 \zeta_5
          + \frac{3664}{3} \zeta_2 \zeta_3
          + \frac{8064}{5} \zeta_2^2
          - \frac{4928}{3} \zeta_2^2 \zeta_3
          - \frac{831904}{945} \zeta_2^3
          + 2 \chi^{g}_{3}
          \bigg] 

       +  C_F^3 n_f   \bigg[2 \chi^{g}_{4} \bigg] \bigg\}\,,
\end{autobreak}
\end{align}

\begin{align}
\begin{autobreak}

   \Delta_{b}^{\rm sv,(4)} =

          \delta(1-z)   \bigg\{

          \frac{1}{2} \g^{b,(1)}_4
          
       +  \frac{d^{abcd}_F d^{abcd}_A}{N_F}    \bigg[ 
            3520 \zeta_2 \zeta_5
          + 128 \zeta_2 \zeta_3
          - 1152 \zeta_2 \zeta_3^2
          - 384 \zeta_2^2
          - \frac{23808}{35} \zeta_2^4
          \bigg] 

       +  n_f   \bigg[ 
          - \frac{2}{3} \g^{b,(2)}_3
          + \frac{d^{abcd}_F d^{abcd}_F}{N_F}    \bigg(
            768 \zeta_2^2
          - 1280 \zeta_2 \zeta_5
          - 256 \zeta_2 \zeta_3
            \bigg)
          \bigg] 

       +  C_A   \bigg[ 
            \frac{11}{3} \g^{b,(2)}_3
          \bigg] 

       +  C_F n_f^3   \bigg[ 
          - \frac{1160}{729}
          + \frac{676}{27} \zeta_5
          + \frac{7028}{729} \zeta_3
          - \frac{1172}{243} \zeta_2
          - \frac{1496}{81} \zeta_2 \zeta_3
          - \frac{766}{81} \zeta_2^2
          \bigg] 

       +  C_F C_A n_f^2   \bigg[ 
            \frac{198202909}{419904}
          - \frac{4216}{27} \zeta_5
          + \frac{625379}{1458} \zeta_3
          + \frac{136}{9} \zeta_3^2
          - \frac{190373}{17496} \zeta_2
          - \frac{9704}{27} \zeta_2 \zeta_3
          + \frac{48433}{270} \zeta_2^2
          - \frac{5018}{315} \zeta_2^3
          \bigg] 

       +  C_F C_A^2 n_f   \bigg[ 
            \frac{22761154}{6561}
          - \frac{422}{9} \zeta_7
          + \frac{864352}{45} \zeta_5
          + \frac{53961407}{1458} \zeta_3
          - \frac{116908}{9} \zeta_3^2
          - \frac{29514571}{2187} \zeta_2
          - \frac{14672}{9} \zeta_2 \zeta_5
          + \frac{248566}{27} \zeta_2 \zeta_3
          + \frac{325069}{180} \zeta_2^2
          - \frac{13312}{45} \zeta_2^2 \zeta_3
          - 118 \zeta_2^3
          \bigg] 

       +  C_F C_A^3   \bigg[ 
          - \frac{634801943}{26244}
          + \frac{2321}{9} \zeta_7
          - \frac{225613}{5} \zeta_5
          - \frac{107597201}{1458} \zeta_3
          + 36421 \zeta_3^2
          + \frac{137480689}{4374} \zeta_2
          + \frac{104456}{9} \zeta_2 \zeta_5
          - \frac{1088924}{27} \zeta_2 \zeta_3
          - 48 \zeta_2 \zeta_3^2
          - \frac{6320831}{810} \zeta_2^2
          + \frac{168256}{45} \zeta_2^2 \zeta_3
          + \frac{42361}{42} \zeta_2^3
          - \frac{20032}{35} \zeta_2^4
          \bigg] 

       +  C_F^2 n_f^2   \bigg[ 
          - \frac{1341097}{5832}
          - \frac{64576}{27} \zeta_5
          - \frac{717916}{729} \zeta_3
          + \frac{27452}{27} \zeta_3^2
          + \frac{360040}{729} \zeta_2
          + \frac{201328}{81} \zeta_2 \zeta_3
          - \frac{261653}{810} \zeta_2^2
          - \frac{111512}{945} \zeta_2^3
          \bigg] 

       +  C_F^2 C_A n_f   \bigg[ 
            \frac{893866105}{236196}
          - 3444 \zeta_7
          - \frac{2961118}{1215} \zeta_5
          - \frac{713487637}{13122} \zeta_3
          + \frac{7565486}{729} \zeta_3^2
          + \frac{79408775}{4374} \zeta_2
          - \frac{161672}{135} \zeta_2 \zeta_5
          - \frac{369562}{9} \zeta_2 \zeta_3
          + \frac{39100019}{14580} \zeta_2^2
          + \frac{2436236}{405} \zeta_2^2 \zeta_3
          + \frac{953776}{315} \zeta_2^3
          \bigg] 

       +  C_F^2 C_A^2   \bigg[ 
            \frac{4432795339}{236196}
          - \frac{14368}{45} \zeta_{5,3}
          + \frac{191542}{9} \zeta_7
          + \frac{7232512}{243} \zeta_5
          + \frac{433756822}{6561} \zeta_3
          - \frac{56606}{9} \zeta_3 \zeta_5
          - \frac{17298667}{729} \zeta_3^2
          - \frac{484431923}{6561} \zeta_2
          + \frac{1589996}{135} \zeta_2 \zeta_5
          + \frac{123525268}{729} \zeta_2 \zeta_3
          - \frac{95108}{27} \zeta_2 \zeta_3^2
          - \frac{52607237}{3645} \zeta_2^2
          - \frac{13402286}{405} \zeta_2^2 \zeta_3
          - \frac{3856}{135} \zeta_2^3
          + \frac{433849}{375} \zeta_2^4
          \bigg] 

       +  C_F^3 n_f   \bigg[ 
            \frac{42292165}{26244}
          + \frac{4134164}{63} \zeta_7
          + \frac{12703004}{405} \zeta_5
          + \frac{3439907}{243} \zeta_3
          - \frac{1498604}{81} \zeta_3^2
          - \frac{713072}{81} \zeta_2
          - \frac{139808}{3} \zeta_2 \zeta_5
          + \frac{301096}{81} \zeta_2 \zeta_3
          - \frac{450404}{1215} \zeta_2^2
          - \frac{232048}{27} \zeta_2^2 \zeta_3
          + \frac{9356848}{2835} \zeta_2^3
          \bigg] 

       +  C_F^3 C_A   \bigg[ 
          - \frac{123478228}{6561}
          - \frac{46048}{15} \zeta_{5,3}
          - \frac{8449718}{21} \zeta_7
          - \frac{84828502}{405} \zeta_5
          + \frac{90494864}{729} \zeta_3
          + \frac{288728}{5} \zeta_3 \zeta_5
          + \frac{7112530}{81} \zeta_3^2
          + \frac{16073350}{243} \zeta_2
          + \frac{2160104}{9} \zeta_2 \zeta_5
          - \frac{5325596}{81} \zeta_2 \zeta_3
          - \frac{423232}{9} \zeta_2 \zeta_3^2
          + \frac{30325238}{1215} \zeta_2^2
          + \frac{5244104}{135} \zeta_2^2 \zeta_3
          - \frac{88860088}{2835} \zeta_2^3
          - \frac{12244}{1125} \zeta_2^4
          \bigg] 

       +  C_F^4   \bigg[ 
            31207
          + \frac{10304}{3} \zeta_{5,3}
          + 34896 \zeta_7
          + \frac{1158368}{15} \zeta_5
          - \frac{472894}{3} \zeta_3
          + \frac{19003904}{45} \zeta_3 \zeta_5
          + \frac{381148}{27} \zeta_3^2
          - 20237 \zeta_2
          + \frac{49056}{5} \zeta_2 \zeta_5
          + \frac{157904}{9} \zeta_2 \zeta_3
          - \frac{3502208}{27} \zeta_2 \zeta_3^2
          - \frac{300664}{15} \zeta_2^2
          + \frac{37936}{5} \zeta_2^2 \zeta_3
          + \frac{4372432}{315} \zeta_2^3
          - \frac{16309024}{525} \zeta_2^4
          \bigg] 
          + n_f \bigg[ \frac{1}{2} \chi^b_1  \bigg] 
          + n_f^0 \bigg[ \frac{1}{2} \chi^b_2 \bigg] 
        
       \bigg\}

       + {\cal D}_1   \bigg\{

          \frac{d^{abcd}_F d^{abcd}_A}{N_F}     \bigg[ 
            \frac{14080}{3} \zeta_5
          + \frac{512}{3} \zeta_3
          - 1536 \zeta_3^2
          - 512 \zeta_2
          - \frac{31744}{35} \zeta_2^3
          \bigg] 

       +  n_f   \frac{d^{abcd}_F d^{abcd}_F}{N_F}   \bigg[ 
          - \frac{5120}{3} \zeta_5
          - \frac{1024}{3} \zeta_3
          + 1024 \zeta_2
          \bigg] 

       +  C_F n_f^3   \bigg[ 
          - \frac{16000}{729}
          + \frac{512}{9} \zeta_3
          + \frac{2560}{27} \zeta_2
          \bigg] 

       +  C_F C_A n_f^2   \bigg[ 
            \frac{232180}{243}
          - \frac{9280}{9} \zeta_3
          - \frac{57920}{27} \zeta_2
          + \frac{2048}{15} \zeta_2^2
          \bigg] 

       +  C_F C_A^2 n_f   \bigg[ 
          - \frac{2285548}{243}
          - \frac{5440}{9} \zeta_5
          + \frac{28000}{3} \zeta_3
          + \frac{407776}{27} \zeta_2
          + 128 \zeta_2 \zeta_3
          - \frac{9856}{5} \zeta_2^2
          \bigg] 

       +  C_F C_A^3   \bigg[ 
            \frac{18081268}{729}
          + \frac{61600}{9} \zeta_5
          - \frac{289312}{9} \zeta_3
          - 64 \zeta_3^2
          - \frac{303328}{9} \zeta_2
          + 2112 \zeta_2 \zeta_3
          + \frac{21824}{3} \zeta_2^2
          - \frac{80128}{105} \zeta_2^3
          \bigg] 

       +  C_F^2 n_f^2   \bigg[ 
            \frac{492040}{729}
          - \frac{350848}{81} \zeta_3
          - \frac{31616}{27} \zeta_2
          + \frac{7168}{135} \zeta_2^2
          \bigg] 

       +  C_F^2 C_A n_f   \bigg[ 
          - \frac{4974308}{729}
          + 512 \zeta_5
          + \frac{4282880}{81} \zeta_3
          + \frac{1484896}{81} \zeta_2
          - 8192 \zeta_2 \zeta_3
          - \frac{432896}{135} \zeta_2^2
          \bigg] 

       +  C_F^2 C_A^2   \bigg[ 
            \frac{9682720}{729}
          - 1344 \zeta_5
          - \frac{12217792}{81} \zeta_3
          + \frac{3008}{3} \zeta_3^2
          - \frac{4981888}{81} \zeta_2
          + \frac{175232}{3} \zeta_2 \zeta_3
          + \frac{2456128}{135} \zeta_2^2
          - \frac{1150592}{315} \zeta_2^3
          \bigg] 

       +  C_F^3 n_f   \bigg[ 
            432
          + \frac{1010944}{9} \zeta_5
          + \frac{434240}{27} \zeta_3
          - \frac{14144}{27} \zeta_2
          - \frac{748544}{9} \zeta_2 \zeta_3
          + \frac{1327232}{135} \zeta_2^2
          \bigg] 

       +  C_F^3 C_A   \bigg[ 
          - \frac{29984}{9}
          - \frac{5730688}{9} \zeta_5
          - \frac{1764608}{27} \zeta_3
          + \frac{224512}{3} \zeta_3^2
          + \frac{624032}{27} \zeta_2
          + \frac{3860864}{9} \zeta_2 \zeta_3
          - \frac{9003968}{135} \zeta_2^2
          + \frac{794624}{63} \zeta_2^3
          \bigg] 

       +  C_F^4   \bigg[ 
            \frac{17248}{3}
          + 13568 \zeta_5
          - 19008 \zeta_3
          + \frac{1148416}{3} \zeta_3^2
          - \frac{21088}{3} \zeta_2
          + 14336 \zeta_2 \zeta_3
          + \frac{18816}{5} \zeta_2^2
          - \frac{4450816}{63} \zeta_2^3
          \bigg]
        \bigg\}

       + {\cal D}_0   \bigg\{
  
         \frac{d^{abcd}_F d^{abcd}_A}{N_F}  \bigg[
          - 2 f^{q}_{4,d^{abcd}_F d^{abcd}_A}
          \bigg] 

       + n_f \frac{d^{abcd}_F d^{abcd}_F}{N_F}   \bigg[ 
            768
          + 4 b^{q}_{4,d^{abcd}_F d^{abcd}_F}
          + \frac{43520}{9} \zeta_5
          + \frac{10624}{9} \zeta_3
          - \frac{2432}{3} \zeta_3^2
          - \frac{9088}{3} \zeta_2
          - 256 \zeta_2 \zeta_3
          + \frac{640}{3} \zeta_2^2
          - \frac{18944}{315} \zeta_2^3
          \bigg] 

       + C_F n_f^3   \bigg[ 
            \frac{10432}{2187}
          - \frac{3680}{81} \zeta_3
          - \frac{3200}{81} \zeta_2
          + \frac{224}{45} \zeta_2^2
          \bigg] 

       + C_F C_A n_f^2   \bigg[ 
          - \frac{898033}{2916}
          + \frac{608}{3} \zeta_5
          + \frac{87280}{81} \zeta_3
          + \frac{293528}{243} \zeta_2
          - \frac{608}{9} \zeta_2 \zeta_3
          - \frac{3488}{15} \zeta_2^2
          \bigg] 

       + C_F C_A^2 n_f   \bigg[ 
            \frac{10761379}{2916}
          - 2 b^{q}_{4,n_f C_F^2 C_A}
          - b^{q}_{4,n_f C_F^3}
          - \frac{1}{12} b^{q}_{4,d^{abcd}_F d^{abcd}_F}
          - \frac{29552}{27} \zeta_5
          - \frac{948884}{81} \zeta_3
          - \frac{9736}{9} \zeta_3^2
          - \frac{2418814}{243} \zeta_2
          + \frac{28064}{9} \zeta_2 \zeta_3
          + \frac{85312}{27} \zeta_2^2
          - \frac{52688}{189} \zeta_2^3
          \bigg] 

       + C_F C_A^3   \bigg[ 
          - \frac{28325071}{2187}
          + \frac{1}{12} f^{q}_{4,d^{abcd}_F d^{abcd}_A}
          + 3400 \zeta_7
          - \frac{49840}{9} \zeta_5
          + \frac{867584}{27} \zeta_3
          - \frac{4664}{3} \zeta_3^2
          + \frac{5761670}{243} \zeta_2
          + 832 \zeta_2 \zeta_5
          - \frac{119624}{9} \zeta_2 \zeta_3
          - \frac{301264}{45} \zeta_2^2
          + 576 \zeta_2^2 \zeta_3
          + \frac{334312}{315} \zeta_2^3
          \bigg] 

       + C_F^2 n_f^2   \bigg[ 
          - \frac{309953}{729}
          + \frac{33056}{9} \zeta_5
          + \frac{124976}{81} \zeta_3
          + \frac{314240}{729} \zeta_2
          - \frac{79360}{27} \zeta_2 \zeta_3
          + \frac{6976}{27} \zeta_2^2
          \bigg] 

       + C_F^2 C_A n_f   \bigg[ 
            \frac{5590667}{1458}
          + 4 b^{q}_{4,n_f C_F^2 C_A}
          - 37616 \zeta_5
          - \frac{1606660}{81} \zeta_3
          + \frac{11728}{3} \zeta_3^2
          - \frac{2713546}{729} \zeta_2
          + \frac{313888}{9} \zeta_2 \zeta_3
          - \frac{1920992}{405} \zeta_2^2
          - \frac{1312}{105} \zeta_2^3
          \bigg] 

       + C_F^2 C_A^2   \bigg[ 
            \frac{1571464}{729}
          + \frac{1005056}{9} \zeta_5
          + \frac{5327504}{81} \zeta_3
          - \frac{88640}{3} \zeta_3^2
          + \frac{6077552}{729} \zeta_2
          + 3072 \zeta_2 \zeta_5
          - \frac{3124160}{27} \zeta_2 \zeta_3
          + \frac{3124352}{405} \zeta_2^2
          + \frac{121888}{15} \zeta_2^2 \zeta_3
          - \frac{34496}{15} \zeta_2^3
          \bigg] 

       + C_F^3 n_f   \bigg[ - \frac{48157}{54}
          + 4 b^{q}_{4,n_f C_F^3}
          - \frac{130624}{3} \zeta_5
          - \frac{19520}{9} \zeta_3
          + \frac{106336}{3} \zeta_3^2
          - \frac{31256}{27} \zeta_2
          + \frac{208256}{9} \zeta_2 \zeta_3
          - \frac{63496}{45} \zeta_2^2
          - \frac{257344}{63} \zeta_2^3
          \bigg] 

       + C_F^3 C_A   \bigg[ 
          - \frac{25856}{27}
          + 274432 \zeta_5
          - \frac{18112}{3} \zeta_3
          - \frac{511840}{3} \zeta_3^2
          - \frac{78080}{27} \zeta_2
          - 73728 \zeta_2 \zeta_5
          - \frac{1275328}{9} \zeta_2 \zeta_3
          + \frac{478336}{45} \zeta_2^2
          + 30400 \zeta_2^2 \zeta_3
          + \frac{406912}{15} \zeta_2^3
          \bigg] 

       + C_F^4   \bigg[ 
            983040 \zeta_7
          - 49152 \zeta_5
          + 4096 \zeta_3
          - 15360 \zeta_3^2
          - 491520 \zeta_2 \zeta_5
          + 32768 \zeta_2 \zeta_3
          - \frac{391168}{5} \zeta_2^2 \zeta_3
          \bigg] 
        \bigg\} \,.

\end{autobreak}
\end{align}

The finite coefficients $\g^{I,(j)}_{i}$ appear from the soft-gluon contributions (See Eq.~(44) of \cite{Ravindran:2005vv} for the details). 
Through the symbols $\chi^q_j$ and $\chi^g_j$, we denote the unknown coefficients of the color factors in four loop form factors for the Drell-Yan and Higgs boson production via gluon fusion, respectively.
For the case of Higgs boson production in bottom quark annihilation, only the $n_f^3$ and $n_f^2$ contributions to the four loop form factor are available in the literature~\cite{Lee:2017mip}. As a result, the unknown coefficients corresponding to $O(n_f)$ and $O(n_f^0)$ color factors are denoted by $\chi^b_1$ and $\chi^b_2$, respectively. Also the symbols $f^q_{4,d_F^{abcd} d_A^{abcd}}$ and $b^q_{4,j}$, where $j = \big\{d_F^{abcd} d_F^{abcd}, n_f C_F^3, n_f C_F^2 C_A, d_F^{abcd} d_A^{abcd}, C_F^2 C_A^2, C_F^3 C_A, C_F^4 \big\}$ are the unknown coefficients of the color factors in four loop soft and collinear anomalous dimensions. 
%
In the above equations, $n_f$ is the number of active light quark flavors, $n_{fv}$ is proportional to the charge weighted sum of the quark flavors and $N_4 = (n_c^2-4)/n_c$ \cite{Gehrmann:2010ue}.
Following ref.~\cite{Henn:2019swt}, we have
\begin{align}
  \frac{d_{A}^{abcd}d_{A}^{abcd}}{N_{A}} &= \frac{n_c^2 (n_c^2+36)}{24} \,,\quad
 \frac{d_{F}^{abcd}d_{A}^{abcd}}{N_{A}} =\frac{n_c (n_c^2+6)}{48} \,,\quad    \frac{d_{F}^{abcd}d_{F}^{abcd}}{N_{A}} = \frac{n_c^4-6 n_c^2+18}{96 n_c^2} \,,\nn\\&
 C_{A} = n_c \,,\quad C_{F} =\frac{n_c^2-1}{2 n_c} \,,\quad N_{A} = n_c^2-1 \,,\quad N_{F} = n_c . 
 \end{align}

%% file: N3LL.tex
\section{Explicit results at N$^3$LL for Drell-Yan and Higgs boson production}
\label{secE:N3LL}
Here, we provide the full explicit result of the resummation constant  ${\overline g}^I_{4}$ given in \eqref{eq:GNbexp} at N$^{3}$LL for Drell-Yan and Higgs boson productions.

\begin{align}
\begin{autobreak}
 {\overline g}^I_{4} =
 
         \frac{1}{(1-\omega)^2}
         \bigg\{ C_R \bigg\{
         
         \frac{\beta_1^3}{\beta_0^5}  \bigg[ 
          - \frac{2}{3} L_\omega^3
          + 2 \omega^2 L_\omega
          + \frac{4}{3} \omega^3
          \bigg] 

       + \frac{\beta_1 \beta_2}{\beta_0^4}   \bigg[
          - 2 L_\omega
          - 2 \omega
          + 4 \omega L_\omega
          + 3 \omega^2
          - 4 \omega^2 L_\omega
          - \frac{8}{3} \omega^3
          \bigg] 

       + \frac{\beta_1^2}{\beta_0^4}  n_f   \bigg[ 
          - \frac{20}{9} L_\omega
          - \frac{20}{9} L_\omega^2
          - \frac{20}{9} \omega
          + \frac{10}{9} \omega^2
          + \frac{40}{27} \omega^3
          \bigg] 

       + \frac{\beta_1^2}{\beta_0^4}  C_A   \bigg[ 
            \frac{134}{9} L_\omega
          + \frac{134}{9} L_\omega^2
          + \frac{134}{9} \omega
          - \frac{67}{9} \omega^2
          - \frac{268}{27} \omega^3
          - 4 \zeta_2 L_\omega
          - 4 \zeta_2 L_\omega^2
          - 4 \zeta_2 \omega
          + 2 \zeta_2 \omega^2
          + \frac{8}{3} \zeta_2 \omega^3
          \bigg] 

       + \frac{\beta_3}{\beta_0^3}    \bigg[ 
            2 L_\omega
          + 2 \omega
          - 4 \omega L_\omega
          - 3 \omega^2
          + 2 \omega^2 L_\omega
          + \frac{4}{3} \omega^3
          \bigg] 

       + \frac{\beta_2}{\beta_0^3}  n_f   \bigg[ 
          - \frac{40}{27} \omega^3
          \bigg] 

       + \frac{\beta_2}{\beta_0^3}  C_A   \bigg[ 
            \frac{268}{27} \omega^3
          - \frac{8}{3} \zeta_2 \omega^3
          \bigg] 

       + \frac{\beta_1}{\beta_0^3} n_f^2   \bigg[ 
            \frac{8}{27} L_\omega
          + \frac{8}{27} \omega
          + \frac{4}{27} \omega^2
          - \frac{16}{81} \omega^3
          \bigg] 

       + \frac{\beta_1}{\beta_0^3}  C_A n_f   \bigg[ 
            \frac{418}{27} L_\omega
          + \frac{418}{27} \omega
          + \frac{209}{27} \omega^2
          - \frac{836}{81} \omega^3
          + \frac{56}{3} \zeta_3 L_\omega
          + \frac{56}{3} \zeta_3 \omega
          + \frac{28}{3} \zeta_3 \omega^2
          - \frac{112}{9} \zeta_3 \omega^3
          - \frac{80}{9} \zeta_2 L_\omega
          - \frac{80}{9} \zeta_2 \omega
          - \frac{40}{9} \zeta_2 \omega^2
          + \frac{160}{27} \zeta_2 \omega^3
          \bigg] 

       + \frac{\beta_1}{\beta_0^3}  C_A^2   \bigg[ 
          - \frac{245}{3} L_\omega
          - \frac{245}{3} \omega
          - \frac{245}{6} \omega^2
          + \frac{490}{9} \omega^3
          - \frac{44}{3} \zeta_3 L_\omega
          - \frac{44}{3} \zeta_3 \omega
          - \frac{22}{3} \zeta_3 \omega^2
          + \frac{88}{9} \zeta_3 \omega^3
          + \frac{536}{9} \zeta_2 L_\omega
          + \frac{536}{9} \zeta_2 \omega
          + \frac{268}{9} \zeta_2 \omega^2
          - \frac{1072}{27} \zeta_2 \omega^3
          - \frac{88}{5} \zeta_2^2 L_\omega
          - \frac{88}{5} \zeta_2^2 \omega
          - \frac{44}{5} \zeta_2^2 \omega^2
          + \frac{176}{15} \zeta_2^2 \omega^3
          \bigg] 

       + \frac{\beta_1}{\beta_0^3}  C_F n_f   \bigg[ 
            \frac{55}{3} L_\omega
          + \frac{55}{3} \omega
          + \frac{55}{6} \omega^2
          - \frac{110}{9} \omega^3
          - 16 \zeta_3 L_\omega
          - 16 \zeta_3 \omega
          - 8 \zeta_3 \omega^2
          + \frac{32}{3} \zeta_3 \omega^3
          \bigg]

       + \frac{1}{\beta_0^2}  n_f^3   \bigg[ 
          - \frac{16}{81} \omega^2
          + \frac{32}{243} \omega^3
          + \frac{32}{27} \zeta_3 \omega^2
          - \frac{64}{81} \zeta_3 \omega^3
          \bigg] 

       + \frac{1}{\beta_0^2}  C_A n_f^2   \bigg[ 
            \frac{923}{162} \omega^2
          - \frac{923}{243} \omega^3
          + \frac{1120}{27} \zeta_3 \omega^2
          - \frac{2240}{81} \zeta_3 \omega^3
          - \frac{304}{81} \zeta_2 \omega^2
          + \frac{608}{243} \zeta_2 \omega^3
          - \frac{112}{15} \zeta_2^2 \omega^2
          + \frac{224}{45} \zeta_2^2 \omega^3
          \bigg] 

       + \frac{1}{\beta_0^2}  C_A^2 n_f   \bigg[ 
          - \frac{24137}{162} \omega^2
          + \frac{24137}{243} \omega^3
          + \frac{1048}{9} \zeta_5 \omega^2
          - \frac{2096}{27} \zeta_5 \omega^3
          - \frac{11552}{27} \zeta_3 \omega^2
          + \frac{23104}{81} \zeta_3 \omega^3
          + \frac{10160}{81} \zeta_2 \omega^2
          - \frac{20320}{243} \zeta_2 \omega^3
          + \frac{224}{3} \zeta_2 \zeta_3 \omega^2
          - \frac{448}{9} \zeta_2 \zeta_3 \omega^3
          - \frac{176}{15} \zeta_2^2 \omega^2
          + \frac{352}{45} \zeta_2^2 \omega^3
          \bigg] 

       + \frac{1}{\beta_0^2}  C_A^3   \bigg[ 
            \frac{42139}{81} \omega^2
          - \frac{84278}{243} \omega^3
          - \frac{1804}{9} \zeta_5 \omega^2
          + \frac{3608}{27} \zeta_5 \omega^3
          + \frac{10472}{27} \zeta_3 \omega^2
          - \frac{20944}{81} \zeta_3 \omega^3
          - 8 \zeta_3^2 \omega^2
          + \frac{16}{3} \zeta_3^2 \omega^3
          - \frac{44200}{81} \zeta_2 \omega^2
          + \frac{88400}{243} \zeta_2 \omega^3
          - \frac{176}{3} \zeta_2 \zeta_3 \omega^2
          + \frac{352}{9} \zeta_2 \zeta_3 \omega^3
          + \frac{1804}{5} \zeta_2^2 \omega^2
          - \frac{3608}{15} \zeta_2^2 \omega^3
          - \frac{10016}{105} \zeta_2^3 \omega^2
          + \frac{20032}{315} \zeta_2^3 \omega^3
          \bigg] 

       + \frac{1}{\beta_0^2}  C_F n_f^2   \bigg[ 
            \frac{1196}{81} \omega^2
          - \frac{2392}{243} \omega^3
          - \frac{320}{9} \zeta_3 \omega^2
          + \frac{640}{27} \zeta_3 \omega^3
          + \frac{32}{5} \zeta_2^2 \omega^2
          - \frac{64}{15} \zeta_2^2 \omega^3
          \bigg] 

       + \frac{1}{\beta_0^2}  C_F C_A n_f   \bigg[ 
          - \frac{17033}{81} \omega^2
          + \frac{34066}{243} \omega^3
          + 80 \zeta_5 \omega^2
          - \frac{160}{3} \zeta_5 \omega^3
          + \frac{1856}{9} \zeta_3 \omega^2
          - \frac{3712}{27} \zeta_3 \omega^3
          + \frac{220}{3} \zeta_2 \omega^2
          - \frac{440}{9} \zeta_2 \omega^3
          - 64 \zeta_2 \zeta_3 \omega^2
          + \frac{128}{3} \zeta_2 \zeta_3 \omega^3
          - \frac{176}{5} \zeta_2^2 \omega^2
          + \frac{352}{15} \zeta_2^2 \omega^3
          \bigg] 

       + \frac{1}{\beta_0^2}  C_F^2 n_f   \bigg[ 
            \frac{286}{9} \omega^2
          - \frac{572}{27} \omega^3
          - 160 \zeta_5 \omega^2
          + \frac{320}{3} \zeta_5 \omega^3
          + \frac{296}{3} \zeta_3 \omega^2
          - \frac{592}{9} \zeta_3 \omega^3
          \bigg] 

       + \frac{\beta_1}{\beta_0^2}   n_f   \bigg[ 
            \frac{112}{27} L_\omega
          + \frac{112}{27} \omega
          - \frac{56}{27} \omega^2
          - \frac{16}{3} \zeta_2 L_\omega
          - \frac{16}{3} \zeta_2 \omega
          + \frac{8}{3} \zeta_2 \omega^2
          \bigg] 

       + \frac{\beta_1}{\beta_0^2}  C_A   \bigg[ 
          - \frac{808}{27} L_\omega
          - \frac{808}{27} \omega
          + \frac{404}{27} \omega^2
          + 28 \zeta_3 L_\omega
          + 28 \zeta_3 \omega
          - 14 \zeta_3 \omega^2
          + \frac{88}{3} \zeta_2 L_\omega
          + \frac{88}{3} \zeta_2 \omega
          - \frac{44}{3} \zeta_2 \omega^2
          \bigg] 

       + \frac{1}{\beta_0}  n_f^2   \bigg[ 
            \frac{1856}{729} \omega
          - \frac{928}{729} \omega^2
          - \frac{160}{27} \zeta_3 \omega
          + \frac{80}{27} \zeta_3 \omega^2
          - \frac{320}{27} \zeta_2 \omega
          + \frac{160}{27} \zeta_2 \omega^2
          \bigg] 

       + \frac{1}{\beta_0}  C_A n_f   \bigg[ 
          - \frac{62626}{729} \omega
          + \frac{31313}{729} \omega^2
          + \frac{1240}{9} \zeta_3 \omega
          - \frac{620}{9} \zeta_3 \omega^2
          + \frac{14696}{81} \zeta_2 \omega
          - \frac{7348}{81} \zeta_2 \omega^2
          - \frac{368}{15} \zeta_2^2 \omega
          + \frac{184}{15} \zeta_2^2 \omega^2
          \bigg] 

       + \frac{1}{\beta_0}  C_A^2   \bigg[ 
            \frac{297029}{729} \omega
          - \frac{297029}{1458} \omega^2
          + 192 \zeta_5 \omega
          - 96 \zeta_5 \omega^2
          - \frac{20072}{27} \zeta_3 \omega
          + \frac{10036}{27} \zeta_3 \omega^2
          - \frac{49112}{81} \zeta_2 \omega
          + \frac{24556}{81} \zeta_2 \omega^2
          + \frac{176}{3} \zeta_2 \zeta_3 \omega
          - \frac{88}{3} \zeta_2 \zeta_3 \omega^2
          + \frac{1496}{15} \zeta_2^2 \omega
          - \frac{748}{15} \zeta_2^2 \omega^2
          \bigg] 

       + \frac{1}{\beta_0}  C_F n_f   \bigg[ 
          - \frac{1711}{27} \omega
          + \frac{1711}{54} \omega^2
          + \frac{304}{9} \zeta_3 \omega
          - \frac{152}{9} \zeta_3 \omega^2
          + 16 \zeta_2 \omega
          - 8 \zeta_2 \omega^2
          + \frac{32}{5} \zeta_2^2 \omega
          - \frac{16}{5} \zeta_2^2 \omega^2
          \bigg] 

       + \frac{\beta_1}{\beta_0}    \bigg[
          - 8 \zeta_2 L_\omega
          \bigg] 

       + n_f   \bigg[ 
          - \frac{160}{9} \zeta_2 \omega
          + \frac{80}{9} \zeta_2 \omega^2
          \bigg] 

       + C_A   \bigg[ 
            \frac{1072}{9} \zeta_2 \omega
          - \frac{536}{9} \zeta_2 \omega^2
          - 32 \zeta_2^2 \omega
          + 16 \zeta_2^2 \omega^2
          \bigg] 

       + \beta_0    \bigg[ 
            \frac{64}{3} \zeta_3 \omega
          - \frac{32}{3} \zeta_3 \omega^2
          \bigg] 
        \bigg\}
        
         + \frac{1}{\beta_0^2} \frac{d^{abcd}_R d^{abcd}_A}{N_R} \bigg[ 
            \frac{1760}{3} \zeta_5 \omega^2
          - \frac{3520}{9} \zeta_5 \omega^3
          + \frac{64}{3} \zeta_3 \omega^2
          - \frac{128}{9} \zeta_3 \omega^3
          - 192 \zeta_3^2 \omega^2
          + 128 \zeta_3^2 \omega^3
          - 64 \zeta_2 \omega^2
          + \frac{128}{3} \zeta_2 \omega^3
          - \frac{3968}{35} \zeta_2^3 \omega^2
          + \frac{7936}{105} \zeta_2^3 \omega^3
          \bigg] 

       + \frac{1}{\beta_0^2} n_f \frac{d^{abcd}_R d^{abcd}_F}{N_R}  \bigg[ 
          - \frac{640}{3} \zeta_5 \omega^2
          + \frac{1280}{9} \zeta_5 \omega^3
          - \frac{128}{3} \zeta_3 \omega^2
          + \frac{256}{9} \zeta_3 \omega^3
          + 128 \zeta_2 \omega^2
          - \frac{256}{3} \zeta_2 \omega^3
          \bigg] 
        
         \bigg\}  
\end{autobreak}
\end{align}         
In the above equation, $L_\omega = \ln(1-\omega)$, $R = A$ for gluons($I=g$) and $R = F$ for quarks($I=q,b$). 
The general results of the resummation constants in terms of universal quantities are presented to N$^3$LL accuracy in the ancillary file supplied with the \arXiv~submission.

%% file: main.bbl
\providecommand{\href}[2]{#2}\begingroup\raggedright\begin{thebibliography}{10}

\bibitem{Brooijmans:2020yij}
G.~Brooijmans et~al., \emph{{Les Houches 2019 Physics at TeV Colliders: New
  Physics Working Group Report}},  in \emph{{11th Les Houches Workshop on
  Physics at TeV Colliders}: {PhysTeV Les Houches}}, 2, 2020,
  \href{https://arxiv.org/abs/2002.12220}{{\ttfamily 2002.12220}}.

\bibitem{Anastasiou:2015ema}
C.~Anastasiou, C.~Duhr, F.~Dulat, F.~Herzog and B.~Mistlberger, \emph{{Higgs
  Boson Gluon-Fusion Production in QCD at Three Loops}},
  \href{https://doi.org/10.1103/PhysRevLett.114.212001}{\emph{Phys. Rev. Lett.}
  {\bfseries 114} (2015) 212001}
  [\href{https://arxiv.org/abs/1503.06056}{{\ttfamily 1503.06056}}].

\bibitem{Mistlberger:2018etf}
B.~Mistlberger, \emph{{Higgs boson production at hadron colliders at N$^{3}$LO
  in QCD}}, \href{https://doi.org/10.1007/JHEP05(2018)028}{\emph{JHEP}
  {\bfseries 05} (2018) 028}
  [\href{https://arxiv.org/abs/1802.00833}{{\ttfamily 1802.00833}}].

\bibitem{Dulat:2018bfe}
F.~Dulat, B.~Mistlberger and A.~Pelloni, \emph{{Precision predictions at
  N$^3$LO for the Higgs boson rapidity distribution at the LHC}},
  \href{https://doi.org/10.1103/PhysRevD.99.034004}{\emph{Phys. Rev.}
  {\bfseries D99} (2019) 034004}
  [\href{https://arxiv.org/abs/1810.09462}{{\ttfamily 1810.09462}}].

\bibitem{Duhr:2019kwi}
C.~Duhr, F.~Dulat and B.~Mistlberger, \emph{{Higgs production in bottom-quark
  fusion to third order in the strong coupling}},
  \href{https://arxiv.org/abs/1904.09990}{{\ttfamily 1904.09990}}.

\bibitem{Mondini:2019gid}
R.~Mondini, M.~Schiavi and C.~Williams, \emph{{N$^{3}$LO predictions for the
  decay of the Higgs boson to bottom quarks}},
  \href{https://doi.org/10.1007/JHEP06(2019)079}{\emph{JHEP} {\bfseries 06}
  (2019) 079} [\href{https://arxiv.org/abs/1904.08960}{{\ttfamily
  1904.08960}}].

\bibitem{Duhr:2020seh}
C.~Duhr, F.~Dulat and B.~Mistlberger, \emph{{The Drell-Yan cross section to
  third order in the strong coupling constant}},
  \href{https://arxiv.org/abs/2001.07717}{{\ttfamily 2001.07717}}.

\bibitem{Duhr:2020kzd}
C.~Duhr, F.~Dulat, V.~Hirschi and B.~Mistlberger, \emph{{Higgs production in
  bottom quark fusion: Matching the 4- and 5-flavour schemes to third order in
  the strong coupling}},  \href{https://arxiv.org/abs/2004.04752}{{\ttfamily
  2004.04752}}.

\bibitem{Chawdhry:2019bji}
H.~A. Chawdhry, M.~L. Czakon, A.~Mitov and R.~Poncelet, \emph{{NNLO QCD
  corrections to three-photon production at the LHC}},
  \href{https://doi.org/10.1007/JHEP02(2020)057}{\emph{JHEP} {\bfseries 02}
  (2020) 057} [\href{https://arxiv.org/abs/1911.00479}{{\ttfamily
  1911.00479}}].

\bibitem{Chen:2019lzz}
L.-B. Chen, H.~T. Li, H.-S. Shao and J.~Wang, \emph{{Higgs boson pair
  production via gluon fusion at N$^3$LO in QCD}},
  \href{https://arxiv.org/abs/1909.06808}{{\ttfamily 1909.06808}}.

\bibitem{Chen:2019fhs}
L.-B. Chen, H.~T. Li, H.-S. Shao and J.~Wang, \emph{{The gluon-fusion
  production of Higgs boson pair: N$^3$LO QCD corrections and top-quark mass
  effects}},  \href{https://arxiv.org/abs/1912.13001}{{\ttfamily 1912.13001}}.

\bibitem{Forslund:2020lnu}
M.~Forslund and N.~Kidonakis, \emph{{Resummation for $2\rightarrow n$ processes
  in single-particle-inclusive kinematics}},
  \href{https://doi.org/10.1103/PhysRevD.102.034006}{\emph{Phys. Rev. D}
  {\bfseries 102} (2020) 034006}
  [\href{https://arxiv.org/abs/2003.09021}{{\ttfamily 2003.09021}}].

\bibitem{deFlorian:2012za}
D.~de~Florian and J.~Mazzitelli, \emph{{A next-to-next-to-leading order
  calculation of soft-virtual cross sections}},
  \href{https://doi.org/10.1007/JHEP12(2012)08,
  10.1007/JHEP12(2012)088}{\emph{JHEP} {\bfseries 12} (2012) 088}
  [\href{https://arxiv.org/abs/1209.0673}{{\ttfamily 1209.0673}}].

\bibitem{Catani:2014uta}
S.~Catani, L.~Cieri, D.~de~Florian, G.~Ferrera and M.~Grazzini,
  \emph{{Threshold resummation at N$^3$LL accuracy and soft-virtual cross
  sections at N$^3$LO}},
  \href{https://doi.org/10.1016/j.nuclphysb.2014.09.012}{\emph{Nucl. Phys.}
  {\bfseries B888} (2014) 75}
  [\href{https://arxiv.org/abs/1405.4827}{{\ttfamily 1405.4827}}].

\bibitem{Korchemsky:1987wg}
G.~Korchemsky and A.~Radyushkin, \emph{{Renormalization of the Wilson Loops
  Beyond the Leading Order}},
  \href{https://doi.org/10.1016/0550-3213(87)90277-X}{\emph{Nucl. Phys. B}
  {\bfseries 283} (1987) 342}.

\bibitem{Moch:2004pa}
S.~Moch, J.~A.~M. Vermaseren and A.~Vogt, \emph{{The Three loop splitting
  functions in QCD: The Nonsinglet case}},
  \href{https://doi.org/10.1016/j.nuclphysb.2004.03.030}{\emph{Nucl. Phys.}
  {\bfseries B688} (2004) 101}
  [\href{https://arxiv.org/abs/hep-ph/0403192}{{\ttfamily hep-ph/0403192}}].

\bibitem{Vogt:2004mw}
A.~Vogt, S.~Moch and J.~A.~M. Vermaseren, \emph{{The Three-loop splitting
  functions in QCD: The Singlet case}},
  \href{https://doi.org/10.1016/j.nuclphysb.2004.04.024}{\emph{Nucl. Phys.}
  {\bfseries B691} (2004) 129}
  [\href{https://arxiv.org/abs/hep-ph/0404111}{{\ttfamily hep-ph/0404111}}].

\bibitem{Henn:2019swt}
J.~M. Henn, G.~P. Korchemsky and B.~Mistlberger, \emph{{The full four-loop cusp
  anomalous dimension in $\mathcal{N}=4$ super Yang-Mills and QCD}},
  \href{https://doi.org/10.1007/JHEP04(2020)018}{\emph{JHEP} {\bfseries 04}
  (2020) 018} [\href{https://arxiv.org/abs/1911.10174}{{\ttfamily
  1911.10174}}].

\bibitem{vonManteuffel:2020vjv}
A.~von Manteuffel, E.~Panzer and R.~M. Schabinger, \emph{{Analytic four-loop
  anomalous dimensions in massless QCD from form factors}},
  \href{https://doi.org/10.1103/PhysRevLett.124.162001}{\emph{Phys. Rev. Lett.}
  {\bfseries 124} (2020) 162001}
  [\href{https://arxiv.org/abs/2002.04617}{{\ttfamily 2002.04617}}].

\bibitem{Ravindran:2004mb}
V.~Ravindran, J.~Smith and W.~L. van Neerven, \emph{{Two-loop corrections to
  Higgs boson production}},
  \href{https://doi.org/10.1016/j.nuclphysb.2004.10.039}{\emph{Nucl. Phys.}
  {\bfseries B704} (2005) 332}
  [\href{https://arxiv.org/abs/hep-ph/0408315}{{\ttfamily hep-ph/0408315}}].

\bibitem{Ravindran:2005vv}
V.~Ravindran, \emph{{On Sudakov and soft resummations in QCD}},
  \href{https://doi.org/10.1016/j.nuclphysb.2006.04.008}{\emph{Nucl. Phys.}
  {\bfseries B746} (2006) 58}
  [\href{https://arxiv.org/abs/hep-ph/0512249}{{\ttfamily hep-ph/0512249}}].

\bibitem{Ravindran:2006cg}
V.~Ravindran, \emph{{Higher-order threshold effects to inclusive processes in
  QCD}}, \href{https://doi.org/10.1016/j.nuclphysb.2006.06.025}{\emph{Nucl.
  Phys.} {\bfseries B752} (2006) 173}
  [\href{https://arxiv.org/abs/hep-ph/0603041}{{\ttfamily hep-ph/0603041}}].

\bibitem{Ahmed:2014cla}
T.~Ahmed, M.~Mahakhud, N.~Rana and V.~Ravindran, \emph{{Drell-Yan Production at
  Threshold to Third Order in QCD}},
  \href{https://doi.org/10.1103/PhysRevLett.113.112002}{\emph{Phys. Rev. Lett.}
  {\bfseries 113} (2014) 112002}
  [\href{https://arxiv.org/abs/1404.0366}{{\ttfamily 1404.0366}}].

\bibitem{Ahmed:2014cha}
T.~Ahmed, N.~Rana and V.~Ravindran, \emph{{Higgs boson production through $b
  \bar b$ annihilation at threshold in N$^3$LO QCD}},
  \href{https://doi.org/10.1007/JHEP10(2014)139}{\emph{JHEP} {\bfseries 10}
  (2014) 139} [\href{https://arxiv.org/abs/1408.0787}{{\ttfamily 1408.0787}}].

\bibitem{Kumar:2014uwa}
M.~C. Kumar, M.~K. Mandal and V.~Ravindran, \emph{{Associated production of
  Higgs boson with vector boson at threshold N$^{3}$LO in QCD}},
  \href{https://doi.org/10.1007/JHEP03(2015)037}{\emph{JHEP} {\bfseries 03}
  (2015) 037} [\href{https://arxiv.org/abs/1412.3357}{{\ttfamily 1412.3357}}].

\bibitem{Ahmed:2015qda}
T.~Ahmed, M.~C. Kumar, P.~Mathews, N.~Rana and V.~Ravindran,
  \emph{{Pseudo-scalar Higgs boson production at threshold N$^3$ LO and N$^3$
  LL QCD}}, \href{https://doi.org/10.1140/epjc/s10052-016-4199-1}{\emph{Eur.
  Phys. J.} {\bfseries C76} (2016) 355}
  [\href{https://arxiv.org/abs/1510.02235}{{\ttfamily 1510.02235}}].

\bibitem{Das:2019bxi}
G.~Das, M.~C. Kumar and K.~Samanta, \emph{{Resummed inclusive cross-section in
  ADD model at N$^3$LL+NNLO}},
  \href{https://arxiv.org/abs/1912.13039}{{\ttfamily 1912.13039}}.

\bibitem{Li:2014bfa}
Y.~Li, A.~von Manteuffel, R.~M. Schabinger and H.~X. Zhu, \emph{{N$^3$LO Higgs
  boson and Drell-Yan production at threshold: The one-loop two-emission
  contribution}}, \href{https://doi.org/10.1103/PhysRevD.90.053006}{\emph{Phys.
  Rev.} {\bfseries D90} (2014) 053006}
  [\href{https://arxiv.org/abs/1404.5839}{{\ttfamily 1404.5839}}].

\bibitem{H:2018hqz}
A.~H. Ajjath, P.~Banerjee, A.~Chakraborty, P.~K. Dhani, P.~Mukherjee, N.~Rana
  et~al., \emph{{Higgs pair production from bottom quark annihilation to NNLO
  in QCD}}, \href{https://doi.org/10.1007/JHEP05(2019)030}{\emph{JHEP}
  {\bfseries 05} (2019) 030}
  [\href{https://arxiv.org/abs/1811.01853}{{\ttfamily 1811.01853}}].

\bibitem{Ahmed:2019udm}
T.~Ahmed, A.~H. Ajjath, L.~Chen, P.~K. Dhani, P.~Mukherjee and V.~Ravindran,
  \emph{{Polarised Amplitudes and Soft-Virtual Cross Sections for $b\bar b
  \rightarrow ZH$ at NNLO in QCD}},
  \href{https://doi.org/10.1007/JHEP01(2020)030}{\emph{JHEP} {\bfseries 01}
  (2020) 030} [\href{https://arxiv.org/abs/1910.06347}{{\ttfamily
  1910.06347}}].

\bibitem{Ahmed:2019syo}
T.~Ahmed, A.~H. Ajjath, L.~Chen, P.~K. Dhani, P.~Mukherjee and V.~Ravindran,
  \emph{{Two-Loop QCD Helicity Amplitudes for Higgs Production Associated with
  a Vector Boson through Bottom Quark Annihilation}},  in \emph{{14th
  International Symposium on Radiative Corrections: Application of Quantum
  Field Theory to Phenomenology (RADCOR 2019) Avignon, France, September 8-13,
  2019}}, 2019, \href{https://arxiv.org/abs/1912.06271}{{\ttfamily
  1912.06271}}.

\bibitem{H:2019nsw}
A.~H. Ajjath, P.~Banerjee, A.~Chakraborty, P.~K. Dhani, P.~Mukherjee, N.~Rana
  et~al., \emph{{NNLO QCD$\oplus$QED corrections to Higgs production in bottom
  quark annihilation}},
  \href{https://doi.org/10.1103/PhysRevD.100.114016}{\emph{Phys. Rev.}
  {\bfseries D100} (2019) 114016}
  [\href{https://arxiv.org/abs/1906.09028}{{\ttfamily 1906.09028}}].

\bibitem{Ajjath:2019vmf}
A.~H. Ajjath, P.~Mukherjee and V.~Ravindran, \emph{{Infrared structure of
  $SU(N)\times U(1)$ gauge theory to three loops}},
  \href{https://arxiv.org/abs/1912.13386}{{\ttfamily 1912.13386}}.

\bibitem{Ravindran:2006bu}
V.~Ravindran, J.~Smith and W.~L. van Neerven, \emph{{QCD threshold corrections
  to di-lepton and Higgs rapidity distributions beyond $N^{2}$ LO}},
  \href{https://doi.org/10.1016/j.nuclphysb.2007.01.005}{\emph{Nucl. Phys.}
  {\bfseries B767} (2007) 100}
  [\href{https://arxiv.org/abs/hep-ph/0608308}{{\ttfamily hep-ph/0608308}}].

\bibitem{Ravindran:2007sv}
V.~Ravindran and J.~Smith, \emph{{Threshold corrections to rapidity
  distributions of $Z$ and $W^\pm$ bosons beyond $N^{2}$ LO at hadron
  colliders}}, \href{https://doi.org/10.1103/PhysRevD.76.114004}{\emph{Phys.
  Rev.} {\bfseries D76} (2007) 114004}
  [\href{https://arxiv.org/abs/0708.1689}{{\ttfamily 0708.1689}}].

\bibitem{Ahmed:2014uya}
T.~Ahmed, M.~K. Mandal, N.~Rana and V.~Ravindran, \emph{{Rapidity Distributions
  in Drell-Yan and Higgs Productions at Threshold to Third Order in QCD}},
  \href{https://doi.org/10.1103/PhysRevLett.113.212003}{\emph{Phys. Rev. Lett.}
  {\bfseries 113} (2014) 212003}
  [\href{https://arxiv.org/abs/1404.6504}{{\ttfamily 1404.6504}}].

\bibitem{Ahmed:2014era}
T.~Ahmed, M.~K. Mandal, N.~Rana and V.~Ravindran, \emph{{Higgs Rapidity
  Distribution in $b {\bar b}$ Annihilation at Threshold in N$^{3}$LO QCD}},
  \href{https://doi.org/10.1007/JHEP02(2015)131}{\emph{JHEP} {\bfseries 02}
  (2015) 131} [\href{https://arxiv.org/abs/1411.5301}{{\ttfamily 1411.5301}}].

\bibitem{Banerjee:2018ozf}
P.~Banerjee, P.~K. Dhani and V.~Ravindran, \emph{{Gluon jet function at three
  loops in QCD}}, \href{https://doi.org/10.1103/PhysRevD.98.094016}{\emph{Phys.
  Rev.} {\bfseries D98} (2018) 094016}
  [\href{https://arxiv.org/abs/1805.02637}{{\ttfamily 1805.02637}}].

\bibitem{Banerjee:2018yrn}
P.~Banerjee, A.~Chakraborty, P.~K. Dhani, V.~Ravindran and S.~Seth,
  \emph{{Second order splitting functions and infrared safe cross sections in $
  \mathcal{N} $ = 4 SYM theory}},
  \href{https://doi.org/10.1007/JHEP04(2019)058}{\emph{JHEP} {\bfseries 04}
  (2019) 058} [\href{https://arxiv.org/abs/1810.07672}{{\ttfamily
  1810.07672}}].

\bibitem{Ahmed:2019hcp}
T.~Ahmed, P.~Banerjee, A.~Chakraborty, P.~K. Dhani, V.~Ravindran and S.~Seth,
  \emph{{Infrared structure of $\mathcal{N}$ = 4 SYM and leading
  transcendentality principle in gauge theory}},
  \href{https://arxiv.org/abs/1912.06929}{{\ttfamily 1912.06929}}.

\bibitem{Anastasiou:2014vaa}
C.~Anastasiou, C.~Duhr, F.~Dulat, E.~Furlan, T.~Gehrmann, F.~Herzog et~al.,
  \emph{{Higgs boson gluon–fusion production at threshold in N$^3$LO QCD}},
  \href{https://doi.org/10.1016/j.physletb.2014.08.067}{\emph{Phys. Lett.}
  {\bfseries B737} (2014) 325}
  [\href{https://arxiv.org/abs/1403.4616}{{\ttfamily 1403.4616}}].

\bibitem{Sterman:1986aj}
G.~F. Sterman, \emph{{Summation of Large Corrections to Short Distance Hadronic
  Cross-Sections}},
  \href{https://doi.org/10.1016/0550-3213(87)90258-6}{\emph{Nucl. Phys. B}
  {\bfseries 281} (1987) 310}.

\bibitem{Catani:1989ne}
S.~Catani and L.~Trentadue, \emph{{Resummation of the QCD Perturbative Series
  for Hard Processes}},
  \href{https://doi.org/10.1016/0550-3213(89)90273-3}{\emph{Nucl. Phys. B}
  {\bfseries 327} (1989) 323}.

\bibitem{Catani:2003zt}
S.~Catani, D.~de~Florian, M.~Grazzini and P.~Nason, \emph{{Soft gluon
  resummation for Higgs boson production at hadron colliders}},
  \href{https://doi.org/10.1088/1126-6708/2003/07/028}{\emph{JHEP} {\bfseries
  07} (2003) 028} [\href{https://arxiv.org/abs/hep-ph/0306211}{{\ttfamily
  hep-ph/0306211}}].

\bibitem{Moch:2005ky}
S.~Moch and A.~Vogt, \emph{{Higher-order soft corrections to lepton pair and
  Higgs boson production}},
  \href{https://doi.org/10.1016/j.physletb.2005.09.061}{\emph{Phys. Lett.}
  {\bfseries B631} (2005) 48}
  [\href{https://arxiv.org/abs/hep-ph/0508265}{{\ttfamily hep-ph/0508265}}].

\bibitem{Bonvini:2014joa}
M.~Bonvini and S.~Marzani, \emph{{Resummed Higgs cross section at N$^{3}$LL}},
  \href{https://doi.org/10.1007/JHEP09(2014)007}{\emph{JHEP} {\bfseries 09}
  (2014) 007} [\href{https://arxiv.org/abs/1405.3654}{{\ttfamily 1405.3654}}].

\bibitem{Bonvini:2016frm}
M.~Bonvini, S.~Marzani, C.~Muselli and L.~Rottoli, \emph{{On the Higgs cross
  section at N$^{3}$LO+N$^{3}$LL and its uncertainty}},
  \href{https://doi.org/10.1007/JHEP08(2016)105}{\emph{JHEP} {\bfseries 08}
  (2016) 105} [\href{https://arxiv.org/abs/1603.08000}{{\ttfamily
  1603.08000}}].

\bibitem{H:2019dcl}
A.~H. Ajjath, A.~Chakraborty, G.~Das, P.~Mukherjee and V.~Ravindran,
  \emph{{Resummed prediction for Higgs boson production through b$
  \overline{\mathrm{b}} $ annihilation at N$^{3}$LL}},
  \href{https://doi.org/10.1007/JHEP11(2019)006}{\emph{JHEP} {\bfseries 11}
  (2019) 006} [\href{https://arxiv.org/abs/1905.03771}{{\ttfamily
  1905.03771}}].

\bibitem{Idilbi:2006dg}
A.~Idilbi, X.-d. Ji and F.~Yuan, \emph{{Resummation of threshold logarithms in
  effective field theory for DIS, Drell-Yan and Higgs production}},
  \href{https://doi.org/10.1016/j.nuclphysb.2006.07.002}{\emph{Nucl. Phys.}
  {\bfseries B753} (2006) 42}
  [\href{https://arxiv.org/abs/hep-ph/0605068}{{\ttfamily hep-ph/0605068}}].

\bibitem{Eynck:2003fn}
T.~O. Eynck, E.~Laenen and L.~Magnea, \emph{{Exponentiation of the Drell-Yan
  cross-section near partonic threshold in the DIS and MS-bar schemes}},
  \href{https://doi.org/10.1088/1126-6708/2003/06/057}{\emph{JHEP} {\bfseries
  06} (2003) 057} [\href{https://arxiv.org/abs/hep-ph/0305179}{{\ttfamily
  hep-ph/0305179}}].

\bibitem{H.:2020ecd}
A.~A. H., G.~Das, M.~C. Kumar, P.~Mukherjee, V.~Ravindran and K.~Samanta,
  \emph{{Resummed Drell-Yan cross-section at N$^3$LL}},
  \href{https://arxiv.org/abs/2001.11377}{{\ttfamily 2001.11377}}.

\bibitem{Moch:2005ba}
S.~Moch, J.~A.~M. Vermaseren and A.~Vogt, \emph{{Higher-order corrections in
  threshold resummation}},
  \href{https://doi.org/10.1016/j.nuclphysb.2005.08.005}{\emph{Nucl. Phys.}
  {\bfseries B726} (2005) 317}
  [\href{https://arxiv.org/abs/hep-ph/0506288}{{\ttfamily hep-ph/0506288}}].

\bibitem{Das:2019btv}
G.~Das, S.-O. Moch and A.~Vogt, \emph{{Soft corrections to inclusive
  deep-inelastic scattering at four loops and beyond}},
  \href{https://arxiv.org/abs/1912.12920}{{\ttfamily 1912.12920}}.

\bibitem{Ahmed:2016otz}
T.~Ahmed, M.~Bonvini, M.~C. Kumar, P.~Mathews, N.~Rana, V.~Ravindran et~al.,
  \emph{{Pseudo-scalar Higgs boson production at N$^3$ LO$_{\text {A}}$ +N$^3$
  LL $'$}}, \href{https://doi.org/10.1140/epjc/s10052-016-4510-1}{\emph{Eur.
  Phys. J.} {\bfseries C76} (2016) 663}
  [\href{https://arxiv.org/abs/1606.00837}{{\ttfamily 1606.00837}}].

\bibitem{Schmidt:2015cea}
T.~Schmidt and M.~Spira, \emph{{Higgs Boson Production via Gluon Fusion:
  Soft-Gluon Resummation including Mass Effects}},
  \href{https://doi.org/10.1103/PhysRevD.93.014022}{\emph{Phys. Rev.}
  {\bfseries D93} (2016) 014022}
  [\href{https://arxiv.org/abs/1509.00195}{{\ttfamily 1509.00195}}].

\bibitem{deFlorian:2007sr}
D.~de~Florian and J.~Zurita, \emph{{Soft-gluon resummation for pseudoscalar
  Higgs boson production at hadron colliders}},
  \href{https://doi.org/10.1016/j.physletb.2007.11.018}{\emph{Phys. Lett.}
  {\bfseries B659} (2008) 813}
  [\href{https://arxiv.org/abs/0711.1916}{{\ttfamily 0711.1916}}].

\bibitem{das:rs}
G.~Das, M.~C. Kumar and K.~Samanta, \emph{{Resummed inclusive cross-section in
  Randall-Sundrum model at NNLO+NNLL}}, .

\bibitem{Catani:1998bh}
S.~Catani, \emph{{The Singular behavior of QCD amplitudes at two loop order}},
  \href{https://doi.org/10.1016/S0370-2693(98)00332-3}{\emph{Phys. Lett.}
  {\bfseries B427} (1998) 161}
  [\href{https://arxiv.org/abs/hep-ph/9802439}{{\ttfamily hep-ph/9802439}}].

\bibitem{Sterman:2002qn}
G.~F. Sterman and M.~E. Tejeda-Yeomans, \emph{{Multiloop amplitudes and
  resummation}},
  \href{https://doi.org/10.1016/S0370-2693(02)03100-3}{\emph{Phys. Lett.}
  {\bfseries B552} (2003) 48}
  [\href{https://arxiv.org/abs/hep-ph/0210130}{{\ttfamily hep-ph/0210130}}].

\bibitem{Moch:2005tm}
S.~Moch, J.~A.~M. Vermaseren and A.~Vogt, \emph{{Three-loop results for quark
  and gluon form-factors}},
  \href{https://doi.org/10.1016/j.physletb.2005.08.067}{\emph{Phys. Lett.}
  {\bfseries B625} (2005) 245}
  [\href{https://arxiv.org/abs/hep-ph/0508055}{{\ttfamily hep-ph/0508055}}].

\bibitem{Aybat:2006wq}
S.~M. Aybat, L.~J. Dixon and G.~F. Sterman, \emph{{The Two-loop anomalous
  dimension matrix for soft gluon exchange}},
  \href{https://doi.org/10.1103/PhysRevLett.97.072001}{\emph{Phys. Rev. Lett.}
  {\bfseries 97} (2006) 072001}
  [\href{https://arxiv.org/abs/hep-ph/0606254}{{\ttfamily hep-ph/0606254}}].

\bibitem{Aybat:2006mz}
S.~M. Aybat, L.~J. Dixon and G.~F. Sterman, \emph{{The Two-loop soft anomalous
  dimension matrix and resummation at next-to-next-to leading pole}},
  \href{https://doi.org/10.1103/PhysRevD.74.074004}{\emph{Phys. Rev.}
  {\bfseries D74} (2006) 074004}
  [\href{https://arxiv.org/abs/hep-ph/0607309}{{\ttfamily hep-ph/0607309}}].

\bibitem{Becher:2009cu}
T.~Becher and M.~Neubert, \emph{{Infrared singularities of scattering
  amplitudes in perturbative QCD}},
  \href{https://doi.org/10.1103/PhysRevLett.102.162001,
  10.1103/PhysRevLett.111.199905}{\emph{Phys. Rev. Lett.} {\bfseries 102}
  (2009) 162001} [\href{https://arxiv.org/abs/0901.0722}{{\ttfamily
  0901.0722}}].

\bibitem{Gardi:2009qi}
E.~Gardi and L.~Magnea, \emph{{Factorization constraints for soft anomalous
  dimensions in QCD scattering amplitudes}},
  \href{https://doi.org/10.1088/1126-6708/2009/03/079}{\emph{JHEP} {\bfseries
  03} (2009) 079} [\href{https://arxiv.org/abs/0901.1091}{{\ttfamily
  0901.1091}}].

\bibitem{Almelid:2015jia}
O.~Almelid, C.~Duhr and E.~Gardi, \emph{{Three-loop corrections to the soft
  anomalous dimension in multileg scattering}},
  \href{https://doi.org/10.1103/PhysRevLett.117.172002}{\emph{Phys. Rev. Lett.}
  {\bfseries 117} (2016) 172002}
  [\href{https://arxiv.org/abs/1507.00047}{{\ttfamily 1507.00047}}].

\bibitem{Almelid:2017qju}
O.~Almelid, C.~Duhr, E.~Gardi, A.~McLeod and C.~D. White, \emph{{Bootstrapping
  the QCD soft anomalous dimension}},
  \href{https://doi.org/10.1007/JHEP09(2017)073}{\emph{JHEP} {\bfseries 09}
  (2017) 073} [\href{https://arxiv.org/abs/1706.10162}{{\ttfamily
  1706.10162}}].

\bibitem{Mueller:1979ih}
A.~H. Mueller, \emph{{On the Asymptotic Behavior of the Sudakov Form-factor}},
  \href{https://doi.org/10.1103/PhysRevD.20.2037}{\emph{Phys. Rev.} {\bfseries
  D20} (1979) 2037}.

\bibitem{Collins:1980ih}
J.~C. Collins, \emph{{Algorithm to Compute Corrections to the Sudakov
  Form-factor}}, \href{https://doi.org/10.1103/PhysRevD.22.1478}{\emph{Phys.
  Rev.} {\bfseries D22} (1980) 1478}.

\bibitem{Sen:1981sd}
A.~Sen, \emph{{Asymptotic Behavior of the Sudakov Form-Factor in QCD}},
  \href{https://doi.org/10.1103/PhysRevD.24.3281}{\emph{Phys. Rev.} {\bfseries
  D24} (1981) 3281}.

\bibitem{Ahmed:2017gyt}
T.~Ahmed, J.~M. Henn and M.~Steinhauser, \emph{{High energy behaviour of form
  factors}}, \href{https://doi.org/10.1007/JHEP06(2017)125}{\emph{JHEP}
  {\bfseries 06} (2017) 125}
  [\href{https://arxiv.org/abs/1704.07846}{{\ttfamily 1704.07846}}].

\bibitem{Moch:2018wjh}
S.~Moch, B.~Ruijl, T.~Ueda, J.~M. Vermaseren and A.~Vogt, \emph{{On quartic
  colour factors in splitting functions and the gluon cusp anomalous
  dimension}},
  \href{https://doi.org/10.1016/j.physletb.2018.06.017}{\emph{Phys. Lett. B}
  {\bfseries 782} (2018) 627}
  [\href{https://arxiv.org/abs/1805.09638}{{\ttfamily 1805.09638}}].

\bibitem{Becher:2006mr}
T.~Becher, M.~Neubert and B.~D. Pecjak, \emph{{Factorization and Momentum-Space
  Resummation in Deep-Inelastic Scattering}},
  \href{https://doi.org/10.1088/1126-6708/2007/01/076}{\emph{JHEP} {\bfseries
  01} (2007) 076} [\href{https://arxiv.org/abs/hep-ph/0607228}{{\ttfamily
  hep-ph/0607228}}].

\bibitem{Becher:2009qa}
T.~Becher and M.~Neubert, \emph{{On the Structure of Infrared Singularities of
  Gauge-Theory Amplitudes}},
  \href{https://doi.org/10.1088/1126-6708/2009/06/081,
  10.1007/JHEP11(2013)024}{\emph{JHEP} {\bfseries 06} (2009) 081}
  [\href{https://arxiv.org/abs/0903.1126}{{\ttfamily 0903.1126}}].

\bibitem{Davies:2016jie}
J.~Davies, A.~Vogt, B.~Ruijl, T.~Ueda and J.~Vermaseren, \emph{{Large-$n_f$
  contributions to the four-loop splitting functions in QCD}},
  \href{https://doi.org/10.1016/j.nuclphysb.2016.12.012}{\emph{Nucl. Phys. B}
  {\bfseries 915} (2017) 335}
  [\href{https://arxiv.org/abs/1610.07477}{{\ttfamily 1610.07477}}].

\bibitem{Das:2020adl}
G.~Das, S.~Moch and A.~Vogt, \emph{{Approximate four-loop QCD corrections to
  the Higgs-boson production cross section}},
  \href{https://arxiv.org/abs/2004.00563}{{\ttfamily 2004.00563}}.

\bibitem{Henn:2016men}
J.~M. Henn, A.~V. Smirnov, V.~A. Smirnov and M.~Steinhauser, \emph{{A planar
  four-loop form factor and cusp anomalous dimension in QCD}},
  \href{https://doi.org/10.1007/JHEP05(2016)066}{\emph{JHEP} {\bfseries 05}
  (2016) 066} [\href{https://arxiv.org/abs/1604.03126}{{\ttfamily
  1604.03126}}].

\bibitem{Lee:2016ixa}
J.~Henn, A.~V. Smirnov, V.~A. Smirnov, M.~Steinhauser and R.~N. Lee,
  \emph{{Four-loop photon quark form factor and cusp anomalous dimension in the
  large-$N_c$ limit of QCD}},
  \href{https://doi.org/10.1007/JHEP03(2017)139}{\emph{JHEP} {\bfseries 03}
  (2017) 139} [\href{https://arxiv.org/abs/1612.04389}{{\ttfamily
  1612.04389}}].

\bibitem{vonManteuffel:2016xki}
A.~von Manteuffel and R.~M. Schabinger, \emph{{Quark and gluon form factors to
  four-loop order in QCD: the $N_f^3$ contributions}},
  \href{https://doi.org/10.1103/PhysRevD.95.034030}{\emph{Phys. Rev. D}
  {\bfseries 95} (2017) 034030}
  [\href{https://arxiv.org/abs/1611.00795}{{\ttfamily 1611.00795}}].

\bibitem{Lee:2017mip}
R.~N. Lee, A.~V. Smirnov, V.~A. Smirnov and M.~Steinhauser, \emph{{The $n_f^2$
  contributions to fermionic four-loop form factors}},
  \href{https://doi.org/10.1103/PhysRevD.96.014008}{\emph{Phys. Rev. D}
  {\bfseries 96} (2017) 014008}
  [\href{https://arxiv.org/abs/1705.06862}{{\ttfamily 1705.06862}}].

\bibitem{vonManteuffel:2019wbj}
A.~von Manteuffel and R.~M. Schabinger, \emph{{Quark and gluon form factors in
  four loop QCD: The $N_f^2$ and $N_{q\gamma} N_f$ contributions}},
  \href{https://doi.org/10.1103/PhysRevD.99.094014}{\emph{Phys. Rev. D}
  {\bfseries 99} (2019) 094014}
  [\href{https://arxiv.org/abs/1902.08208}{{\ttfamily 1902.08208}}].

\bibitem{Gehrmann:2010ue}
T.~Gehrmann, E.~W.~N. Glover, T.~Huber, N.~Ikizlerli and C.~Studerus,
  \emph{{Calculation of the quark and gluon form factors to three loops in
  QCD}}, \href{https://doi.org/10.1007/JHEP06(2010)094}{\emph{JHEP} {\bfseries
  06} (2010) 094} [\href{https://arxiv.org/abs/1004.3653}{{\ttfamily
  1004.3653}}].

\bibitem{Gehrmann:2010tu}
T.~Gehrmann, E.~Glover, T.~Huber, N.~Ikizlerli and C.~Studerus, \emph{{The
  quark and gluon form factors to three loops in QCD through to
  O(eps\textasciicircum{}2)}},
  \href{https://doi.org/10.1007/JHEP11(2010)102}{\emph{JHEP} {\bfseries 11}
  (2010) 102} [\href{https://arxiv.org/abs/1010.4478}{{\ttfamily 1010.4478}}].

\bibitem{Gehrmann:2014vha}
T.~Gehrmann and D.~Kara, \emph{{The $Hb\bar{b}$ form factor to three loops in
  QCD}}, \href{https://doi.org/10.1007/JHEP09(2014)174}{\emph{JHEP} {\bfseries
  09} (2014) 174} [\href{https://arxiv.org/abs/1407.8114}{{\ttfamily
  1407.8114}}].

\bibitem{Vermaseren:2000nd}
J.~A.~M. Vermaseren, \emph{{New features of FORM}},
  \href{https://arxiv.org/abs/math-ph/0010025}{{\ttfamily math-ph/0010025}}.

\bibitem{Ruijl:2017dtg}
B.~Ruijl, T.~Ueda and J.~Vermaseren, \emph{{FORM version 4.2}},
  \href{https://arxiv.org/abs/1707.06453}{{\ttfamily 1707.06453}}.

\end{thebibliography}\endgroup
